\documentclass[%
 reprint,
 amsmath,amssymb,
 aps,
]{revtex4-2}

\usepackage{graphicx}% Include figure files
\usepackage{dcolumn}% Align table columns on decimal point
\usepackage{bm}% bold math
\usepackage{subcaption}
\usepackage{graphicx}
\usepackage{comment}
\usepackage{graphics}
\usepackage{xcolor}
\usepackage{colortbl}
\usepackage{xpatch}

\makeatletter

    \def\CT@@do@color{%
      \global\let\CT@do@color\relax
            \@tempdima\wd\z@
            \advance\@tempdima\@tempdimb
            \advance\@tempdima\@tempdimc
    \advance\@tempdimb\tabcolsep
    \advance\@tempdimc\tabcolsep
    \advance\@tempdima2\tabcolsep
            \kern-\@tempdimb
            \leaders\vrule
                    \hskip\@tempdima\@plus  1fill
            \kern-\@tempdimc
            \hskip-\wd\z@ \@plus -1fill }
    \makeatother

\begin{document}

\preprint{APS/123-QED}

\title{Asymmetric games on networks: towards an Ising-model representation}% Force line breaks with \\
%\thanks{A footnote to the article title}%

\author{A. D. Correia}
\email{a.duartecorreia@uu.nl}%Lines break automatically or can be forced with \\
\author{L. L. Leestmaker} 
\author{H. T. C. Stoof}%
\affiliation%
{Institute for Theoretical Physics and Center for Complex System Studies, \\
Utrecht University, The Netherlands
}%

\author{J. Broere}
\affiliation{Department of Sociology/ICS and Center for Complex System Studies, \\ Utrecht University, The Netherlands}

\date{\today}% It is always \today, today,
             %  but any date may be explicitly specified

\begin{abstract}
We here study the Battle of the Sexes game, a textbook case of asymmetric games, on small networks. Due to the conflicting preferences of the players, analytical approaches are scarce and most often update strategies are employed in numerical simulations of repeated games on networks until convergence is reached. As a result, correlations between the choices of the players emerge. Our approach is to study these correlations with a generalized Ising model. First, we show that these correlations emerge in simulations and can have an Ising-like form. Then, using the response strategy framework, we describe how the actions of the players can bring the network into a steady configuration, starting from an out-of-equilibrium one. We obtain these configurations using game-theoretic tools, and describe the results using Ising parameters. We exhaust the two-player case, giving a detailed account of all the equilibrium possibilities. Going to three players, we generalize the Ising model and compare the equilibrium solutions of three representative types of networks. We find that players that are not directly linked retain a degree of correlation that is proportional to their initial correlation. We also find that the local network structure is the most relevant for small values of the magnetic field and the interaction strength of the Ising model. Finally, we conclude that certain parameters of the equilibrium states are network independent, which opens up the possibility of an analytical description of asymmetric games played on networks.
\\

\textbf{Keywords:} asymmetric games, correlated games, Ising model, games on networks.
\end{abstract}

%\keywords{Suggested keywords}%Use showkeys class option if keyword
                              %display desired
\maketitle

%\tableofcontents

\section{\label{sec:level1}Introduction}
Many different systems rely on coordination processes in order to be successful, whether it is cars in traffic, organizing a company or biological systems\cite{schelling1980strategy,crawford2008power,mehta1994focal}. Understanding coordination processes is often of vital importance. Coordination can be complicated when different actors have different preferences for which option they are willing to coordinate on. In game theory, these situations can be formalized by means of asymmetric (Battle of the Sexes) games. These games reflect situations in which different parties have different preferences for an outcome, but still want to reach coordination. They can, for instance, be used to describe cooperative negotiations, taking place at the political, economical or interpersonal levels\cite{banks1992battle}. Recent studies have shown that networks structures can have an important effect on the behavior of players in 2 $\times$ 2  asymmetric  games\cite{broere2017network,broere2019experimental,hernandez2013heterogeneous,hernandez2017equilibrium, mazzoli2017equilibria}. These studies often rely on numerical simulations of repeated games until convergence is reached.  
We are here interested in giving an analytical description of the equilibrium states of the Battle of the Sexes game played on networks. The goal of this study is to be able to reproduce analytically efficiently the results obtained in numerical simulations in an efficient way\cite{broere2017network,mazzoli2017equilibria}. 

In general, 2 $\times$ 2 games are defined by their characteristic payoff structure\cite{fudenberg1991game}. Given a specified game, every player receives a reward that depends not only on their own actions, but also on those of their opponents. In one-shot games, the players are independent, and each player chooses the option they think yields the highest payoff. This often poses a challenge towards reaching optimal states\cite{nash1950equilibrium}. Because of the uncertainty about what the opponents will do, players might reach an equilibrium that is either not optimal or else even fail to reach an equilibrium altogether. In repeated games, players play the same 2 $\times$ 2 game multiple times with the same opponent. In each round, players are informed about the choice that the other player took in the round before. This information can be accounted for in the next round. Because in repeated games players know the choice of the other players in the previous round, correlations are introduced in the choice behavior.

More recently (repeated) game theory got merged with network theory\cite{boccaletti2006complex,tanimoto2015fundamentals}. The merging of network theory and evolutionary game theory has a long history\cite{szabo2007evolutionary,roca2009effect}, with considerable contributions made using statistical physics approaches\cite{hauert2005game,thurner2009statistical, perc2017statistical}.
 For symmetric games on networks, where all nodes are identical, the only parameters per node are the number of available choices and the degree of connectednes of the nodes. In this case, statistical treatments and analytical predictions of equilibrium configurations exist\cite{lieberman2005evolutionary,allen2017evolutionary}. This is, however, harder to obtain for asymmetric games, because, although each player still has the same number of choices, there are now different kinds of players, each one characterized by her preference for a certain choice\cite{raiffa1957games}. When distributed on a network, there is an unfolding of games being played due to these different preferences: players can either be connected with a player with the same preference (where they play a Pure Coordination (PC) game), or with one with a different preference (where they play a Battle of the Sexes game)\cite{broere2017network}. Taking into account that each node can experience both types of connections if it has a degree of connectedness higher than one, makes studying the convergence of this game on networks extremely challenging.

 Different approaches are employed to simulate games on networks. With \textit{one-shot games}, each player has one chance of making a move, and hence results similar to the uncorrelated (pure and mixed strategy) Nash equilibria are obtained, carrying over the same conflicting issues due to lack of communication. With \textit{iterated games}, by contrast, players continue playing the game repeatedly, and they can adjust their strategies using information about the outcomes from previous rounds. This introduces a form of asynchronous communication, taking the form of learning rules, or \textit{update strategies}. In this setting, a player updates her beliefs about what she should play in the next iteration of the game, based on information from the previous one, until the criteria for convergence are reached \cite{roca2009effect}. The net effect of these learning rules is that the actions of the players become increasingly correlated, to the point of reaching homogeneity. By introducing update strategies, players are oftentimes able to coordinate on one favorable strategy, typically also dependent on the network structure itself\cite{santos2006evolutionary, buskens2016effects}. 

There are two main types of update strategies\cite{roca2009effect}. One type is \textit{proportional imitation}, in which the players mimic the actions of others if these have higher payoffs than their own, usually with a difference above a certain smooth threshold instantiated by, for instance, a Fermi distribution function\cite{cheng2012behavior}. Another one is \textit{best response}, where the players change their action towards the action which would have given them a higher payoff in the previous round\cite{cimini2015dynamics, buskens2016effects}. Best response strategies lead to convergence on a state where the players cannot improve their payoff anymore by changing strategy\cite{anderlini1996path}, which is the defining characteristic of Nash equilibria. For asymmetric games on networks, this is particularly useful since Nash equilibria are hard to reach without any form of communication\cite{banks1992battle}. Hernandez \textit{et al.}\cite{hernandez2013heterogeneous,hernandez2013heterogeneous,hernandez2017equilibrium} predict that, when making choices towards a stable point on the payoff, and if an equilibrium cannot be reached where everyone plays in accordance to their preference, the least connected players and those with a minority identity will then most likely play against their preference. 

Broere \textit{et al.}\cite{broere2017network} implement a numerical simulation for the Battle of the Sexes game, which is asymmetric, using a reinforcement learning that converges to a myopic best response in a wide range of networks, showing that degree centrality is one of the most important predictors of the outcomes of nodes, leading to homogeneity within and heterogeneity between different clusters, which is in accordance with the results of Hernandez \textit{et al.} Follow up experiments with human subjects confirm the numerical predictions\cite{broere2019experimental}. 

 Our proposal here is to look at the underlying correlations that are introduced via the update strategies, corresponding to a form of communication between the players that is necessary to resolve the preference conflict\cite{banks1992battle}, and study them using the parameters of a generalized Ising model. Doing this, we seek to understand how the correlations of outcomes evolve towards equilibrium, using the Ising parameters as an intuitive mapping. We expect that starting with small networks and shifting perspective in this way, we will be able to find consistent patterns that potentiate the extension of this analysis to larger networks.

Obtaining information about the evolving correlations on a network is of vital importance to understand these results from an analytical standpoint. To retrieve the final correlated configurations on a network, one can examine what correlations could be imposed \textit{a priori} on the system as a whole, such that the results have identical final configurations. We contribute to the field, firstly, by showing that simulations using update strategies show a convergence that is explainable with underlying correlations and, secondly, by proposing a combined game-theory and statistical physics method to study the equilibria obtained in games on networks, from the point of view of the underlying correlations introduced by the update rules in the game and of the effects that the choices of the players have in bringing the system to a equilibrium. A translation of these correlations to an Ising model both demonstrates how the simulation equilibria are correlated and reveals interesting emergent properties of the results. Our approach is thus able to take into account the particularities of the network structure and the information shared therein. 

This paper is organized as follows. In Sec. \ref{simulations} we simulate two types of networks that contain the Battle of the Sexes game using the methodology of Broere \textit{et al.}\cite{broere2017network}. We express the final results using an Ising model and show that the total magnetization after convergence supports the existance of underlying correlations. In Sec. \ref{twoplayergames} we lay out the method for assessing equilibria in the presence of correlations, namely for the two-player Battle of the Sexes and Pure Coordination games. We translate the correlations before and after equilibrium to an Ising model, studying how the parameters of the Ising model renormalize under the influence of the actions of the players. In Sec. \ref{threeplayergames} we extend our analysis to three-player networks where one of the players has a different preference than the other two. We apply several renormalization schemes to initial correlations. We then compare different networks based on which correlations represent equilibria under a certain renormalization scheme. For this comparison, we use the parameters of a generalized Ising model, making some symmetry assumptions. In Sec. \ref{discussion} we discuss results such as the emergence of asymptotic behavior common to all networks, as well as some limitations and possible future directions. We conclude in Sec. \ref{conclusion} with a summary of this work and an overview of its expected impact.

\section{Asymmetric games and Ising Model}

In this section we introduce the Nash equilibria as solutions of games where the players act independently from each other, applying the slope analysis as a method for calculating them. We explore this for games with and without correlations, and show how the probabilities of the outcomes can be interpreted in terms of an Ising model.

\subsection{Games without correlations}

A standard assumption in game theory is that players are independent from one another, implying that they cannot communicate directly. Thus, a key challenge towards reaching convergence to an optimal state is that the payoffs obtained by the players depend not only on their individual choices, but also on those of their adversaries.
One possible solution consists in finding a stable point on the payoff. This occurs if each player chooses a strategy for which their individual payoff does not increase if they change strategy. Once every player settles in such a strategy, we obtain a Nash equilibrium\cite{nash1950equilibrium}. There are two types of strategies that can become Nash equilibria: pure strategies, that occur when the players choose to make a move with probability equal to $1$; and mixed strategies, occurring when the players make a move with a certain probability between $0$ and $1$, that makes them indifferent to their adversaries’ choices. Pure strategies are thus particular and extreme cases of mixed strategies.

\begin{table*}[t]
\centering
\begin{subtable}{.3\textwidth}
\centering
\begin{tabular}{c|cc}
 & C & D \\ \hline
C & ($1$, $s_1$) & ($0$, $0$)  \\ 
D & ($0$, $0$) & ($s_2$, $1$)
\end{tabular}
\subcaption{BoS game.}
\label{bosgame}
\end{subtable}
\begin{subtable}{.3\textwidth}
\centering
\begin{tabular}{c|cc}
 & C & D \\ 
 \hline
C & ($1$, $1$) & ($0$, $0$)  \\ 
D & ($0$, $0$) & ($s_1$, $s_1$) 
\end{tabular}
\subcaption{PC game, preference for C.}
\label{pcgameC}
\end{subtable}
\begin{subtable}{.3\textwidth}
\centering
\begin{tabular}{c|cc}
 & C & D \\ \hline
C & ($s_2$, $s_2$) & ($0$, $0$)  \\ 
D & ($0$, $0$) & ($1$, $1$)
\end{tabular}
\subcaption{PC game, preference for D.}
\label{pcgameD}
\end{subtable}
\caption{Payoff tables of the three two-player games occurring in the network structure, with $s\in [0,1]$.}
\label{payoffs3}
\end{table*}

\subsubsection{Slope Analysis}

By this token, Nash equilibria can be thought of as extreme points on the players’ payoff functions, considering the variables to be the probabilities of taking the available actions. A Nash equilibrium is reached when the payoff of each player $i$ does not improve when she changes her probability $P^i_{\mu_i}$ of playing $\mu_i$. This can happen for two reasons: either the payoff function decreases around this probability value, or this is already at the limit of the probability interval $[0,1]$ and therefore it cannot be changed further. 
Taking this view, an analysis of the slopes of the payoff function was introduced by Correia and Stoof\cite{correia2019nash} as a straightforward way to calculate all Nash equilibria. The general idea is that the slope, or coefficient, of a player's payoff function that is associated with the probability $P^i_{\mu_i}$, with which that same player chooses a particular action, already has all the necessary information to calculate the equilibria. On the one hand, the sign of the slope indicates the type of equilibrium. A negative or positive slope means that the equilibrium will be pure, as the probabilities are truncated at $0$ and $1$, respectively, and a zero-valued slope indicates that it will be mixed. On the other hand, a coefficient depends on the probabilities of the actions taken by the adversaries, which have pure or mixed values, in accordance with their own coefficients. When all these conditions are made consistent and evaluated simultaneously, the outcomes are precisely the Nash equilibria.

The probabilities obtained using this method are still independent from each other. For instance, in a game with two players $i={1,2}$ that play $\mu_1,\mu_{2} \in \{C,D\}$, the probability that the final configuration is $(\mu_1,\mu_{2})=(C,C)$ is the product of the probability that player $1$ plays C and the probability that player $2$ also plays C. 
It is due to this independence that the Nash equilibria can be either a sub-optimal solution, or extremely hard to reach. In symmetric games, which describe many social, economical, political and biological phenomena, everyone has the same incentives to cooperate or defect. However, it can still be the case that the ideal results are difficult to achieve without communication. For the Snowdrift game, a Nash equilibrium requires that the identical players take different actions, while for the Prisoner's Dilemma game the players settle on a Nash equilibrium that is not optimal. Other phenomena are better modeled by asymmetric games such as the Battle of the Sexes game\cite{binmore2007playing}, where the conflicting preferences of the players makes the Nash equilibrium hard to reach as well.

\subsubsection{Battle of the Sexes}

The BoS game is a two-player game (Table \ref{bosgame}) where the players find it difficult to agree on the equilibria, because they have distinct individual drives. An illustration of a situation that prompts such a setup involves two colleagues at a conference who would like to attend together one of two talks offered in a parallel session. To their exasperation, they cannot find each other and their phones are not working properly. One colleague has a preference for attending the talk on \textbf{\textit{C}}rystal Spectroscopy, while the other has a preference to hear about \textbf{\textit{D}}-Branes. While both have an incentive to go to the same talk, their conflicting preferences don't allow them to agree on which one without communicating.

Coordinating on the same option, with both players either choosing C or D, forms the two pure-strategy Nash equilibria, making it a coordination game. Additionally, the game has a mixed-strategy Nash equilibrium, where player $1$ plays C or D with probabilities $P^1_C=1/(1+s)$ and $P^1_D=1-P^1_C$ respectively, and player $2$ plays C or D with probabilities $P^2_C=P^1_D$ and $P^2_D=P^1_C$. While the players can in principle agree on this strategy, it leads to a sub-optimal payoff of $s/(1+s)$ for each player, lower than the worst possible outcome at the pure equilibria for any player, which is $s$. On a network, connected identical players will play an additional symmetric coordination game, the Pure Coordination (PC) game, with payoff tables given in Tables \ref{pcgameC} or \ref{pcgameD}, depending on their preference. These games have the same pure and mixed Nash equilibria as the BoS game, but coordinating on their preference is the only rational option that realizes the highest possible payoff of $1$.

\subsection{Correlations in games and Ising model}

Games of two players can be extended with correlations. Correlations have been introduced in game theory to formalize the possibility of communication between the players. This is done by introducing a probability distribution $p_{\mu_1,\mu_2}$, that describes the final joint outcome of the players\cite{aumann1987correlated}. 
To model the statistics of the outcome of games, we look at a one-dimensional Ising model, describing magnetic interactions between interacting particles. The actions C and D are mapped to the spin states ``up" and ``down" of particles with spin $1/2$, such that $\mu_i \in \{-1,1\}$. Particles can couple to an external magnetic field $B_i$, specific to each particle. Additionally, the strength of their interaction is determined by the symmetric interaction strength $J$. 

The probability of finding two particles in a certain configuration of states is given by the Boltzmann distribution
\begin{equation} \label{initialproabilities}
    p_{\mu_1, \mu_{2}}=\frac{1}{Z}e^{-H_{\mu_1, \mu_{2}}},
\end{equation} with $H_{\mu_1 \mu_{2}}$ the Ising Hamiltonian, 

\begin{equation}\label{hamiltonian}
    H_{\mu_1 ,\mu_{2}}=-J\mu_1\mu_{2} - B_1\mu_1 - B_2 \mu_2,
\end{equation} and $Z$ the partition function,
\begin{equation}
Z=\sum_{\mu_1,\mu_2} e^{-H_{\mu_1 \mu_{2}}}.
\end{equation} Because the Nash equilibria of the original game generates uncorrelated joint probabilities, their translation to an Ising model only requires the individual magnetic fields $B_i$ as parameters, such that 
\begin{equation}\label{probs}
    P^i_{\mu_i}=\frac{e^{B_i \mu_i}}{Z_i},
\end{equation} with the individual partition function $Z_i=\sum_{\mu_i} e^{B_i \mu_i}$. There is no need for an interaction strength in that case, which conveys the independent character of the players' actions. 
In this uncorrelated case, where $J=0$, the joint probability distribution is thus equal to the multiplication of the probabilities that the individual spins are in that state, $p_{\mu_1,\mu_2}=P^1_{\mu_1}P^2_{\mu_2}$. The probability of states of the uncorrelated landscape in Eq. \ref{hamiltonian} is equal to the probabilities of the uncorrelated (one-shot) mixed strategy obtained from the payoff Table \ref{bosgame} if the parameters relate in the following manner
\begin{equation}\label{B1}
B_1=\textrm{ln}\sqrt{\frac{1}{s_1}},
\end{equation}
and
\begin{equation}\label{B2}
B_2=-\textrm{ln}\sqrt{\frac{1}{s_2}}.
\end{equation}
For example, assuming that $s_1=s_2=1/2$, the mixed strategy solution in this situation is for player 1 to play D two third of the times and C one third of the times and for player 2 to play D two third of the times and C one third of the times. Therefore, the probabilities of the different strategy combinations are $2/9$ for (C,C), $4/9$ for (C,D), $1/9$ for (D,C) and $2/9$ for (D,D).  The probability of being in the ($1,1$) spin state, for instance, can now too be calculated by filling in Eq. (\ref{initialproabilities}), resulting in $p_{1,1}=2/9$ for the parameters
$J=0$ and $B_1=-B_2=\textrm{ln}\sqrt{2}$.
Filling in Eq. (\ref{initialproabilities}) for the other states with the same parameters yields $p_{1,-1}=4/9$, $p_{-1,1}=1/9$ and $p_{-1,-1}=2/9$, as designed.

It is in bypassing this independence that simulations of games are able to reach a better convergence in outcomes, as the update strategies introduce a non-zero interaction strength J \footnote{Other proposals have similarly suggested the use of an Ising model to this end, but with an identification between the payoff structure and the energy landscape that governs such interactions\cite{benjamin2019triggers}, solely to recover the standard Nash equilibria. However, as Eq.(\ref{probs}) shows, that relationship is unnecessary to describe those equilibria. In our model, by contrast, the payoff structure is imposed on arbitrary correlations via the slope analysis, which are only then mapped to the appropriate Ising parameters.}. 

\section{Correlations in Simulations}\label{simulations}

In this section we use the Ising model to calculate analytical expressions for small networks of interacting binary variables, where we vary the interaction coefficient $J$ in order to represent the repeated interactions of the network games. By further generalizing the Ising model, it can be parametrized to represent the asymmetric situation of the Battle of the Sexes for ring graphs. We then use a computational method and the Ising model to study the equilibrium behavior for small networks.

\subsection{Ising model with a node-dependent magnetic field}

The Battle of the Sexes is an asymmetric game, which means that the same choice combination can have a different payoff for different players. Therefore we need to accomodate the different preferences of the players in the game. We can do this by allowing the system variables to vary between the particles. 
Consider a one-dimensional ring graph $G$ with $N$ nodes. Each node $i$ is interacting with two neighbors on a line with periodic boundary conditions, such that $\mu_{N+1}=\mu_1$.
The total energy of the system for a certain spin configuration $\{\mu_i\}$ can be defined as

\begin{equation}\label{main2}
	H_{\{\mu_i\}} = -\sum_{i=1}^{N}J \mu_i \mu_{i+1}-\sum_{i=1}^{N}B_i\mu_i,
\end{equation}
where for simplicity we assume that $J$ is the same between all nodes, but $B_i$ can take on different values for different $i$ due to their different preferences.
The probability of a certain network configuration is 
\begin{equation}
	p_{\{\mu_i\}}=\frac{e^{-H_{\{\mu_i\}}}}{Z_N}.
\end{equation} The partition function $Z_N$ is obtained by summing over all possible spin configurations: 
	\begin{equation} \label{steps}
	\begin{aligned}
	& Z_N  =  \sum_{\{\mu_i\}} \textrm{exp}\left\{\sum_{i=1}^N\left[J\mu_i\mu_{i+1}+\frac{B_i}{2}\mu_i+\frac{B_{i+1}}{2}\mu_{i+1}\right]\right\} \\
	&= \sum_{\{\mu_i\}} \prod_{i=1}^N \textrm{exp}{\left[J\mu_i\mu_{i+1}+\frac{B_i}{2}\mu_i+\frac{B_{i+1}}{2}\mu_{i+1}\right]} \\
	&= \sum_{\{\mu_i\}} \prod_{i=1}^N T_{\mu_i,\mu_{i+1}},
	\end{aligned}
	\end{equation} with 
	\begin{equation}
	T_{\mu_i,\mu_i+1}=\textrm{exp}\left[J\mu_i\mu_{i+1}+\frac{B_i}{2}\mu_i+\frac{B_{i+1}}{2}\mu_{i+1}\right],
	\end{equation} the elements of the transfer matrix  as
	\begin{equation}\label{bostransfer}
	\textrm{\textbf{T}}= \begin{pmatrix} T_{1,1} & T_{1,-1}\\ T_{-1,1} & T_{-1,-1} \end{pmatrix}.
	\end{equation}The transfer matrix represents the Boltzmann weights of all possible states of $\mu_i$ and $\mu_{i+1}$.

As already described, when the Battle of the Sexes is played on a network interaction structure, three types of interactions can occur, making available the three payoff structures in Table \ref{payoffs3}. Because all three situations need to be accommodated for, we need to define three different transfer matrices. We define the transfer matrix in Eq. (\ref{bostransfer}) for interactions of opposite preferences. For interactions with equal preferences for one type, we define
	\begin{equation} \label{T1}
	\textrm{\textbf{\~{T}}}= \begin{pmatrix} \tilde{T}_{1,1} & \tilde{T}_{1,-1}\\ \tilde{T}_{-1,1} & \tilde{T}_{-1,-1} \end{pmatrix}
	\end{equation}
	and, for the other type,
	\begin{equation} \label{T2}
	\textrm{\textbf{\^{T}}}= \begin{pmatrix} \hat{T}_{1,1} & \hat{T}_{1,-1}\\ \hat{T}_{-1,1} & \hat{T}_{-1,-1} \end{pmatrix}.
	\end{equation} The partition function can now be written as the product of the trace of the $n$-th power of the different transfer matrices $\textbf{T}$, $\textbf{\~{T}}$ and $\textbf{\^{T}}$. Schematically,
	\begin{equation}
	Z= \textrm{tr}[\textrm{\textbf{T}}^{n}\textrm{\textbf{\~{T}}}^{\tilde{n}}\textrm{\textbf{\^{T}}}^{\hat{n}}],
	\label{Trans}
	\end{equation}
	where $n$ refers to number of links with transfer matrix $\textbf{T}$, $\tilde{n}$ refers to number of links with transfer matrix $\textbf{\~{T}}$,  $\hat{n}$ refers to number of links with transfer matrix $\textbf{\^{T}}$ and the total number of links is $n+\tilde{n}+\hat{n}=N$. In Eq. (\ref{Trans}) we have not explicitly denoted the order of variations of the transfer matrices, but only the total number. However, each transfer matrix represents a link between nodes, so the order of the transfer matrices is dependent on the ordering of the different types of nodes in the network. The ordering is important because the transfer matrices do not commute. In the next section we describe the different orderings of nodes that correspond with different orderings of the transfer matrices.

The magnetizations of the network are given by
\begin{equation}
M_i= \frac{\partial }{\partial B_i}\textrm{ln}Z,
\end{equation}
whereas the total magnetization per spin is given by
\begin{equation}\label{avmag}
m= \frac{1}{N}\sum_iM_i.
\end{equation}

\noindent Note that the relationship between the magnetic field and payoff parameters, given by the relations in Eqs. (\ref{B1}) and (\ref{B2}) in all transfer matrices, correspond to the uncorrelated solution of the BoS game, and not of the PC game. Thus, in principle, the values of $B_i$ in Eqs. (\ref{T1}) and (\ref{T2}) should be given by the mixed-strategy solution of their respective PC games. However, this would introduce a conflict in nodes that take part in both a BoS game and a PC game. As such, we use the magnetic fields as Eqs. (\ref{B1}) and (\ref{B2}) for all nodes of either network, anticipating that these values of magnetic field in the presence of a PC game are only possible if $J\neq 0$.

\subsection{Simulation of ring networks}

\begin{figure}[!tbp]
		\centering
		\captionsetup[subfigure]{labelformat=empty}
		\begin{subfigure}[b]{0.24\textwidth}
			\includegraphics[width=\textwidth]{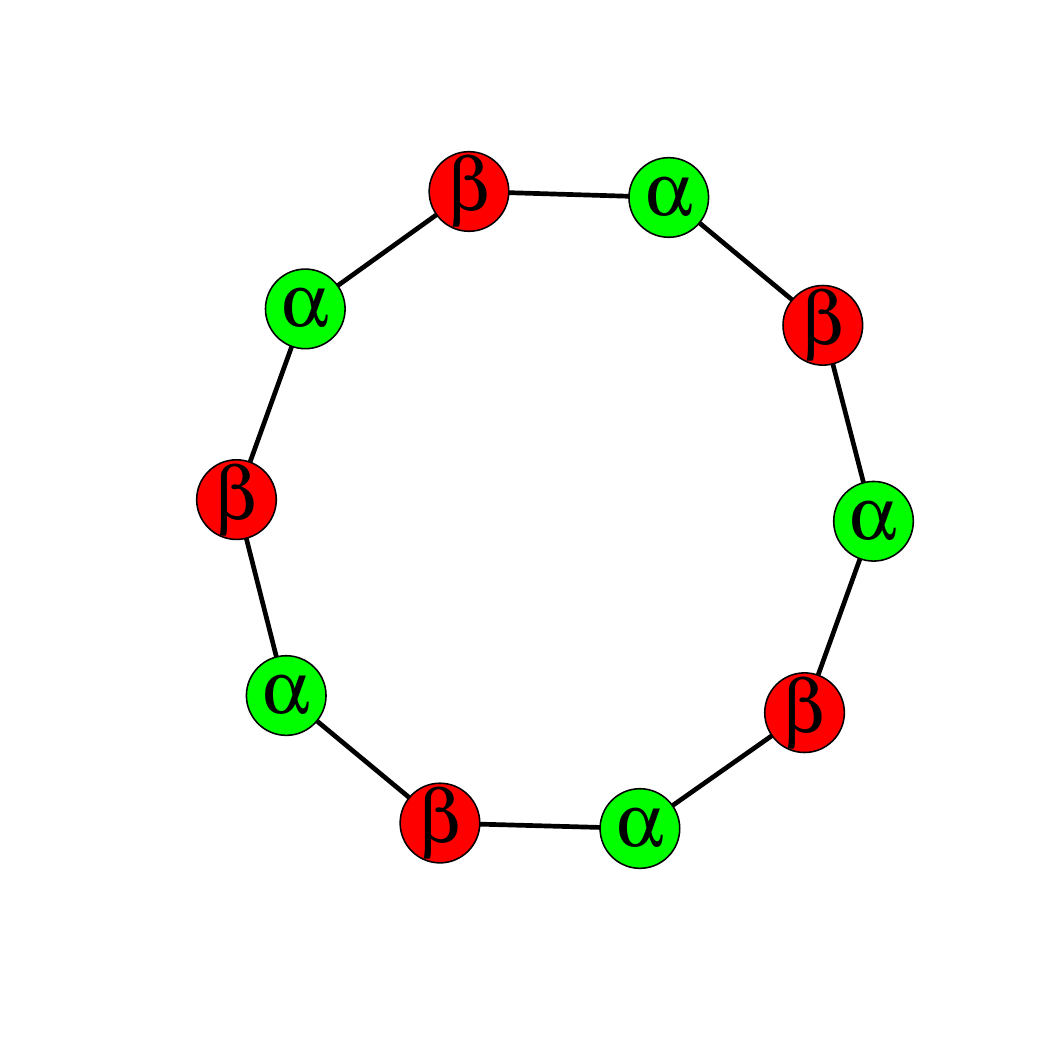}
			\caption{Network 1}
			\label{Network 1}
		\end{subfigure}
		\hfill
		% \begin{minipage}[b]{0.4\textwidth}
		%   \includegraphics[width=\textwidth]{"Variance Man Vrouw".pdf}
		%   \caption{Variance in the network}
		% \end{minipage}\\ 
		%\centering
		\begin{subfigure}[b]{0.24\textwidth}
			\includegraphics[width=\textwidth]{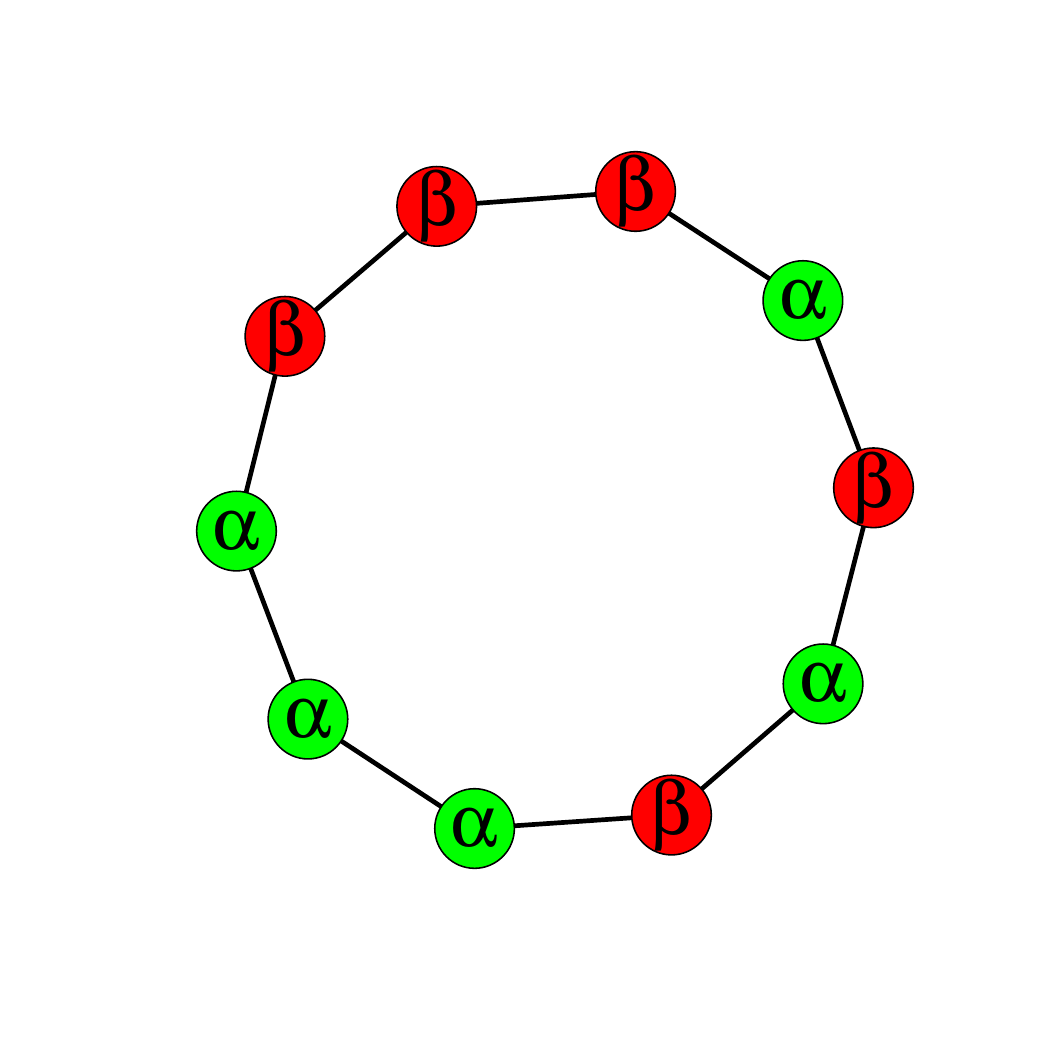}
			\caption{Network 2}
			\label{Network 2}
		\end{subfigure}
		\caption{Ring networks with N=10. The colors indicate the preferences, with the green nodes ($\alpha$) preferring the C equilibrium and the red nodes ($\beta$) preferring the D equilibrium. Note that in Network 1 only BoS games are being played at every edge.}
		\label{Networks}
	\end{figure}

We compare the results of the Ising model to the results of a computational study for two different ring networks described in Fig. \ref{Networks}, which illustrate two different distributions of preference, with the green nodes denoting a preference for C and the red nodes a preference for D. While Fig. \ref{Network 1} represents a completely integrated network where the nodes only interact with nodes with a different preference, by contrast Fig. \ref{Network 2} represents a partly segregated network, where some nodes participate in both a PC and a BoS game.

To simulate the game-theoretic equilibrium for each network, we focus on the computational model described in Broere \textit{et al.}\cite{broere2017network}. In this study, nodes play iterated $2 \times 2$ games with their neighbors. Each node gets a preference assigned before the first round of the game (C or D) that determines whether the node is a row or a column player from the corresponding game in Table \ref{payoffs3}. Every round the nodes choose between C or D and the decision of the node is played against all its neighboring nodes. In the first round, the nodes play their preferred behavior with probability one. After every round, the nodes update their probability of playing C or D by means of reinforcement learning  \cite{cimini2014learning,ezaki2017reinforcement}. The probability of choosing either C or D is updated using a best response strategy, towards what would have been the best choice in the previous round.

	\begin{figure*}[t]
	\centering
	\begin{subfigure}[b]{0.8\textwidth}
		\includegraphics[width=1\linewidth]{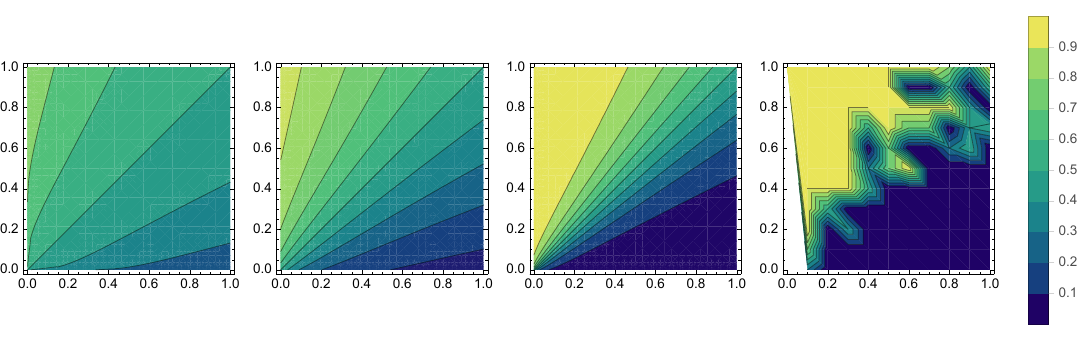}
		\caption{Network 1}
		\label{figS1} 
	\end{subfigure}

	\begin{subfigure}[b]{0.8\textwidth}
		\includegraphics[width=1\linewidth]{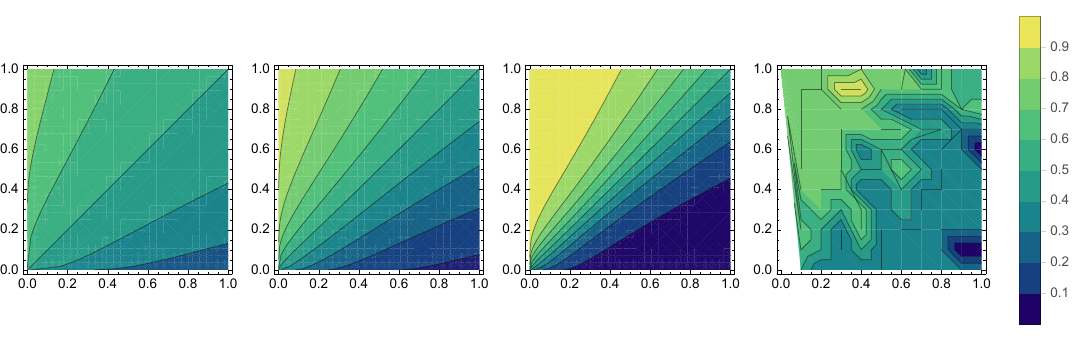}
		\caption{Network 2}
		\label{figS3} 
	\end{subfigure}

	\caption[Probability of choosing $\alpha$]{Probability of choosing C, where the $x$-axis represents the values of $B_1$ in terms of $s_1$ and $y$-axis the value of $B_2$ in terms of $s_2$, according to Eqs. (\ref{B1}) and (\ref{B2}). On the left the plotted values for $J=0$, in the second plot the values for $J=0.5$, in the third plot the values for $J=1$ and on the right the simulation results.}
	\label{figS}
\end{figure*}

We calculate the average magnetization for the different networks, as a function of the magnetic field values, according to Eq. (\ref{avmag}). Per network, the results are obtained for three values of $J$, namely $0$, $0.5$, and $1$, chosen to be representative values of variability. As we want to compare this with the statistics of the equilibrium configurations from the simulations, in Figs. \ref{figS} the average magnetization is plotted as a function of the magnetic field values but with the $x$ and $y$ axis scaled in terms of $s_1$ and $s_2$, following Eqs. (\ref{B1}) and (\ref{B2}), respectively. On the $z$-axis, the average magnetization per spin is scaled to the probability of choosing C, such that an average magnetization of $-1$ means that the probability of playing C is $0$, and an average magnetization of $1$ means that that same probability of $1$. The first three plots are the results from the Ising model with first results for $J=0$, second for $J=0.5$, and third for $J=1$, and the rightmost plot are the results of the simulation of the iterated game-theoretic model. The results for $J=0$ are equal to the uncorrelated mixed equilibrium solution for the whole network.

We first notice that the results are much noisier for the simulations, even with higher resolution and an increased number of simulations. In general, the trend of the simulations is comparable to that of the Ising model, where the simulation results seem to be mostly in line with the Ising model results with $J=1$. This indicates that there is a high correlation in behavior on the networks, and as expected more so for Network 2 than for Network 1. However, we would expect that for Network 1 a clear trend would emerge, but this in fact is not the case. While the simulations confirm the existence of correlations between the players in equilibrium, introduced by the update strategies, they also indicate that a more detailed analytical model is necessary to explain how the correlations evolve in terms of the players decisions. How we can start to think of a generalized Ising model that takes this into account will be the subject of the next two sections.

\section{Response strategies of two-player Games}\label{twoplayergames}

In this section we introduce correlations  analitically in a two-player game, and show how the actions of the players alone can bring a system into equilibrium by renormalizing the underlying correlations. By mapping the correlations to an Ising model, we map out which actions the players are allowed to take towards this goal, given their initial out-of-equilibrium correlations.

\subsection{Correlated Equilibrium}

To evaluate whether it is in equilibrium, the correlated probability distribution in Eq. (\ref{initialproabilities}) is reified with a \textit{correlation device}. This concept was introduced by Aumman\cite{aumann1987correlated} as a concrete way to describe how each player only receives partial information about the entire correlations landscape, remaining blind to the marginals of her opponents. This device draws a final configuration from the probability distribution, and informs each player about what they should play in order for the system to reach that global configuration. A correlated equilibrium, which is a Nash equilibrium, is reached when the players, upon receiving an advice, do not have any incentive to choose any strategy other than the recommended one, while assuming that the opponents do not deviate from their own received instructions. If one of the players wants to deviate from the recommendation, then a correlated equilibrium is not a viable solution. In that case, stability is achieved if the players discard the correlations and resort to the uncorrelated Nash equilibria. 

Correlations were used early on in evolutionary game theory\cite{cripps1991correlated}, but, in spite of some exceptions\cite{metzger2018evolution}, to the best of our knowledge they have not become a standard tool. This can be in part due to the rigidity of the correlated equilibrium definition. This definition of equilibrium, while assuming the existence of underlying correlations in the system, fails to account for the fact that the reactions of the players to an initially out-of-equilibrium probability distribution \textit{can}, if done with enough precision, take the network to a configuration that is in correlated equilibrium. In fact, this is what happens on iterated games with best response strategies, as the actions of the players bring the network from an out-of-equilibrium state to one in which the players do not benefit anymore by changing their strategy. 

\begin{figure*}[t]
\centering
\begin{subfigure}{.45\textwidth}
\centering
\includegraphics[width=.8\linewidth]{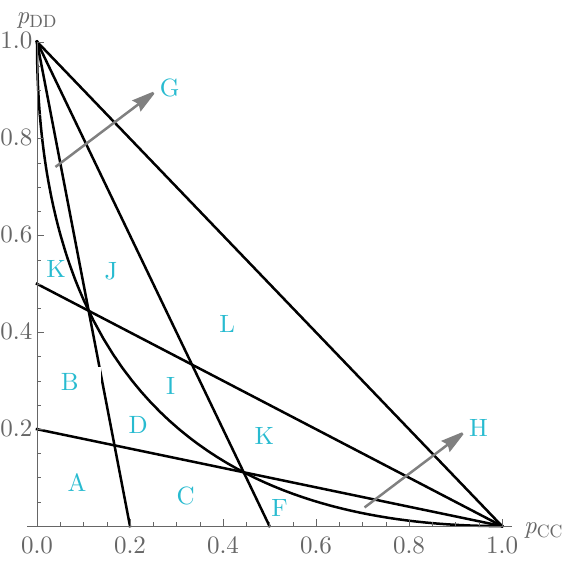}
\caption{}
\label{fig:PC2_combined}
\end{subfigure}
\begin{subfigure}{.45\textwidth}
\centering
\includegraphics[width=.8\linewidth]{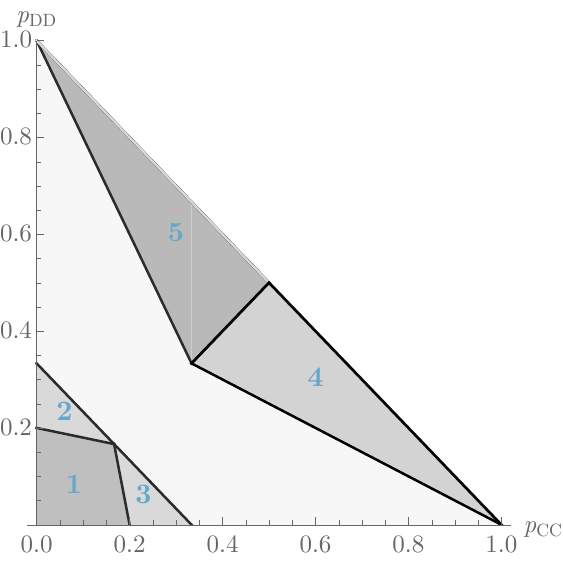}
\caption{}
\label{fig:maxpayoff}
\end{subfigure}
\caption{Characterization of the plane of symmetric correlations. (a) Regions in the $p_{CD}=p_{DC}$ plane of initial correlations corresponding to specific combinations of response strategies that are equilibrium solutions of the two player games, for $s=1/2$. The determinant of the correlation device is positive above the curved line. (b) Areas where a specific response strategy in equilibrium gives a higher payoff than $s=1/2$ in the BoS game. In region 1, both $(1,1,0,0)$ and $(0,0,1,1)$ are equilibria and are equally good solutions for both players. In region 2, $(0,0,1,1)$ is the best solution, which happens for $(1,1,0,0)$ in region 3. In regions 4 and 5 both strategies $(0,0,0,0)$ and $(1,1,1,1)$ are equilibria, but in region 4 player $1$ has a preference for strategy $(1,1,1,1)$, while player $2$ prefers strategy $(0,0,0,0)$, contrasting with region 5, where the preferences of the players are reversed.}
\label{symmetricfig}
\end{figure*}

\subsubsection{Response Probabilities}

Correia and Stoof\cite{correia2019nash} introduce a framework to calculate exactly how the players can act to achieve an equilibrium in this setting. Introducing the \textit{initial} and \textit{renormalized} correlation devices, respectively $p_{\mu'_1,\mu'_2}$ and $p^R_{\mu_1,\mu_2}$, a player $i$ follows probabilistically an instruction $\mu'_i$ given by an initial device. Therefore, instead of strategies consisting of probability distributions over the actions of the original game, now the strategies are distributions over the actions of \textit{following the advice to play $\mu'_i$} and \textit{not following the advice to play $\mu'_i$}, respectively given by the probabilities $P_{F\mu'_i}$ and $P_{NF\mu'_i}=1-P_{F\mu'_i}$, forming the \textit{response strategies}, where the subscripts $F$ and $NF$ qualify the probabilities, indicating whether they are probabilities to F\textit{ollow} or N\textit{ot} F\textit{ollow} the suggestions of the correlation device. The probability of a change in strategy is thus defined in terms of the response strategies as \begin{equation}\label{transtion}
    P^i_{\mu_i\leftarrow\mu_i'}=\delta_{\mu_i\mu_i'}P^i_{F\mu'_i}+(1-\delta_{\mu_i\mu_i'})P^i_{NF{\mu'_i}}.
\end{equation} 

The initial correlation device does not need to be in correlated equilibrium, as a range of other equilibria in these new strategies is possible. An effective, renormalized device is reached as a result of the reactions of the players to the initial one, with

\begin{equation}\label{2pprobs}
    p^R_{\mu_1\mu_2}=\sum_{\mu_1'\mu_2'}P^1_{\mu_1\leftarrow\mu'_1}P^2_{\mu_2\leftarrow\mu_2'}p_{\mu'_1\mu'_2}.
\end{equation}

A correlated equilibrium is reached if a player acts as if she was following an advice to play $\mu_i$, given by this renormalized device. This defines a new game, the \textit{correlated game}. The expected payoff of a player $i$, computed as

\begin{align}\label{payoff}
    & \langle u^i \rangle = \sum_{-i} \sum_{\mu_i\mu_{-i}}u^{}_{\mu_i\mu_{-i}}p^{R}_{\mu_i\mu_{-i}}, 
\end{align} where $-i$ indicates all the opponents of player $i$, which, following the usual game-theoretic convention, corresponds for two players to $i=1$ and $-i=2$. To reach correlated equilibrium, the payoff must be such that it does not increase with a unilateral change in the player's probabilities to follow the new instructions. The response strategies that obey this criterium are the solutions of the correlated game. Players are still independent, since they only react to the external correlations, which means that the response strategy solutions are Nash equilibria as well. Therefore, here too we can use the slope analysis, but now with the variables being the response probabilities to follow and not follow, and with the payoff function depending on the initial correlations as well, such that

\begin{align}\label{slopepayoff}
\langle u^i \rangle =  C^{i}_C P^i_{FC}+C^{i}_D P^i_{FD}+C^{i},
\end{align} with $C^{i}_{\mu_i}$ the slopes of the payoff of player $i$ associated with variable $P^i_{F\mu_i}$, and $C^{i}$ a constant. Note that these coefficients depend on the type and number of interactions that player $i$ has with her neighbors.

We apply this method to the BoS and PC games using $s_1=s_2=1/2$ as in the ring-network simulations, for simplicity we rename $s_1$ and $s_2$ as $s$ from here on out. We study the symmetric section of the space of correlations 
\begin{equation}
p_{CD}=p_{DC}=\frac{1}{2}(1-p_{CC}-p_{DD}),
\end{equation} which allows for a comparison with previous work on symmetric games. However, differently from these games, on the BoS game the players and their strategies are not interchangeable. Hence, we perform a slope analysis on a full set of response probabilities $(P^1_{FC}, P^1_{FD}, P^2_{FC}, P^2_{FD})$. First, we calculate which initial correlations fulfill the slope conditions for player $1$, given the strategy of player $2$, followed by the same calculation with the players reversed. Then, we check which initial correlations are simultaneously solutions for both players, indicating that the chosen response strategy is a Nash equilibrium of the correlated game for those initial correlations. Of the strategies that are equilibria, we compare the expected payoff of the players.

\subsubsection{Equilibrium Strategies}

From the $81$ possible response strategy solutions that arise taking into account $2$ players, $2$ response probabilities for each player and $3$ types of solutions for each response probability, we use symmetry arguments to restrict our study to $23$ strategies that lead to relevant Nash equilibria (see App. \ref{combinatorics}). Each of them is a solution for only a limited range of initial correlation probabilities. Assuming a player $1$ with preference for playing C and a player $2$ with preference for playing D in the BoS game, and two other identical players playing a PC game with C preference, we schematize in Fig. \ref{fig:PC2_combined} the regions where different response strategies are equilibria of these games. The sign of the determinant $D=p_{CC}p_{DD}-p_{DC}p_{CD}$ relates to the sign of the correlations. Positive correlations are beneficial for coordination games, such as BoS and PC, since the players have the highest payoff when both adopt the same strategy. Therefore, negative correlations elicit opposite responses from the players to guarantee that the renormalized correlations are positive again.  

\begin{table*}[t]
    \centering
    \renewcommand{\arraystretch}{1.5}
    \setlength{\tabcolsep}{3pt}
    \begin{tabular}{c|c c c c c c c c c c c c | c}
        $\left(P^1_{FC}, P^1_{FD}, P^2_{FC}, P^2_{FD}\right)$ & A & B & C & D & E & F & G & H & I & J & K & L & Class\\
        \hline
        (1,0,1,0) & $\times$ $\circ$ \cellcolor[HTML]{FF6B00} & $\times$ $\circ$ \cellcolor[HTML]{FF6B00}& $\times$ $\circ$ \cellcolor[HTML]{FF6B00}& $\times$ $\circ$ \cellcolor[HTML]{FF6B00}& $\times$ $\circ$ \cellcolor[HTML]{FF6B00}& $\times$ $\circ$ \cellcolor[HTML]{FF6B00}& $\times$ $\circ$ \cellcolor[HTML]{FF6B00}& $\times$ $\circ$ \cellcolor[HTML]{FF6B00}& $\times$ $\circ$ \cellcolor[HTML]{FF6B00}& $\times$ $\circ$\cellcolor[HTML]{FF6B00} & $\times$ $\circ$\cellcolor[HTML]{FF6B00} & $\times$ $\circ$\cellcolor[HTML]{FF6B00} & $1$\\
        (0,1,0,1) & $\times$ $\circ$ \cellcolor[HTML]{FF6B00}& $\times$ $\circ$ \cellcolor[HTML]{FF6B00}& $\times$ $\circ$ \cellcolor[HTML]{FF6B00}& $\times$ $\circ$ \cellcolor[HTML]{FF6B00}& $\times$ $\circ$ \cellcolor[HTML]{FF6B00}& $\times$ $\circ$ \cellcolor[HTML]{FF6B00} & $\times$ $\circ$ \cellcolor[HTML]{FF6B00}& $\times$ $\circ$ \cellcolor[HTML]{FF6B00}& $\times$ $\circ$ \cellcolor[HTML]{FF6B00}& $\times$ $\circ$ \cellcolor[HTML]{FF6B00}& $\times$ $\circ$ \cellcolor[HTML]{FF6B00} & $\times$ $\circ$ \cellcolor[HTML]{FF6B00}&\\
        ($P^{1*}_{FC}$,$P^{1*}_{FD}$,$P^{2*}_{FC}$,$P^{2*}_{FD}$) & $\times$ $\circ$ \cellcolor[HTML]{FF6B00}& $\times$ $\circ$ \cellcolor[HTML]{FF6B00} & $\times$ $\circ$ \cellcolor[HTML]{FF6B00}& $\times$ $\circ$ \cellcolor[HTML]{FF6B00}& $\times$ $\circ$ \cellcolor[HTML]{FF6B00}& $\times$ $\circ$ \cellcolor[HTML]{FF6B00}& $\times$ $\circ$ \cellcolor[HTML]{FF6B00}& $\times$ $\circ$ \cellcolor[HTML]{FF6B00}& $\times$ $\circ$ \cellcolor[HTML]{FF6B00}& $\times$ $\circ$ \cellcolor[HTML]{FF6B00}& $\times$ $\circ$\cellcolor[HTML]{FF6B00} &$\times$ $\circ$ \cellcolor[HTML]{FF6B00}&  \\ 
        & &  & & & & & & & &  & & & \\
        \hline
        (1,1,1,1) & \cellcolor[HTML]{64A8CC} & \cellcolor[HTML]{64A8CC} & \cellcolor[HTML]{64A8CC} & \cellcolor[HTML]{64A8CC} & \cellcolor[HTML]{64A8CC} & \cellcolor[HTML]{64A8CC} & \cellcolor[HTML]{64A8CC} & \cellcolor[HTML]{64A8CC} & \cellcolor[HTML]{64A8CC} & $\times$ \cellcolor[HTML]{E9CA76}& \cellcolor[HTML]{64A8CC} & $\times$ $\circ$ \cellcolor[HTML]{FF6B00}& $2$ \\
        (0,0,0,0) & \cellcolor[HTML]{64A8CC} & \cellcolor[HTML]{64A8CC} & \cellcolor[HTML]{64A8CC} & \cellcolor[HTML]{64A8CC} & \cellcolor[HTML]{64A8CC} & \cellcolor[HTML]{64A8CC} & \cellcolor[HTML]{64A8CC} & \cellcolor[HTML]{64A8CC} & \cellcolor[HTML]{64A8CC} & \cellcolor[HTML]{64A8CC}  & $\times$ \cellcolor[HTML]{E9CA76}& $\times$ $\circ$ \cellcolor[HTML]{FF6B00} & \\
        (1,1,0,0) & $\times$ $\circ$ \cellcolor[HTML]{FF6B00}& \cellcolor[HTML]{64A8CC} & $\circ$ \cellcolor[HTML]{DC8F01}& \cellcolor[HTML]{64A8CC} &\cellcolor[HTML]{64A8CC}  & \cellcolor[HTML]{64A8CC} & \cellcolor[HTML]{64A8CC} & \cellcolor[HTML]{64A8CC} & \cellcolor[HTML]{64A8CC} & \cellcolor[HTML]{64A8CC} & \cellcolor[HTML]{64A8CC} & \cellcolor[HTML]{64A8CC} & \\
        (0,0,1,1) & $\times$ $\circ$ \cellcolor[HTML]{FF6B00}& $\circ$ \cellcolor[HTML]{DC8F01}& \cellcolor[HTML]{64A8CC} & \cellcolor[HTML]{64A8CC} & \cellcolor[HTML]{64A8CC} & \cellcolor[HTML]{64A8CC} & \cellcolor[HTML]{64A8CC} & \cellcolor[HTML]{64A8CC} & \cellcolor[HTML]{64A8CC} & \cellcolor[HTML]{64A8CC} & \cellcolor[HTML]{64A8CC} & \cellcolor[HTML]{64A8CC} & \\
         & &  & & & & & & & &  & & & \\
         \hline
        (1,$P^{1*}_{FD}$,1,$P^{2*}_{FD}$) & \cellcolor[HTML]{64A8CC} & \cellcolor[HTML]{64A8CC}  & \cellcolor[HTML]{64A8CC} & \cellcolor[HTML]{64A8CC} & \cellcolor[HTML]{64A8CC} & \cellcolor[HTML]{64A8CC} & \cellcolor[HTML]{64A8CC} & $\times$ $\circ$ \cellcolor[HTML]{FF6B00}& \cellcolor[HTML]{64A8CC} & $\times$ $\circ$ \cellcolor[HTML]{FF6B00}& \cellcolor[HTML]{64A8CC} & $\times$ \cellcolor[HTML]{E9CA76} & $3$\\
        
        ($P^{1*}_{FC}$,0,$P^{2*}_{FC}$,0) & \cellcolor[HTML]{64A8CC} & \cellcolor[HTML]{64A8CC} & \cellcolor[HTML]{64A8CC} & \cellcolor[HTML]{64A8CC} & \cellcolor[HTML]{64A8CC} & \cellcolor[HTML]{64A8CC} & $\times$ \cellcolor[HTML]{E9CA76}& $\times$ \cellcolor[HTML]{E9CA76}& $\circ$ \cellcolor[HTML]{DC8F01} & $\times$ \cellcolor[HTML]{E9CA76}& $\times$ $\circ$\cellcolor[HTML]{FF6B00} & $\times$ $\circ$\cellcolor[HTML]{FF6B00} &  \\
        ($P^{1*}_{FC}$,1,$P^{2*}_{FC}$,1) & \cellcolor[HTML]{64A8CC} & \cellcolor[HTML]{64A8CC} & \cellcolor[HTML]{64A8CC} & \cellcolor[HTML]{64A8CC} &\cellcolor[HTML]{64A8CC}  & \cellcolor[HTML]{64A8CC} & \cellcolor[HTML]{64A8CC} & \cellcolor[HTML]{64A8CC} & $\times$ $\circ$\cellcolor[HTML]{FF6B00} & $\times$ \cellcolor[HTML]{E9CA76}& $\circ$ \cellcolor[HTML]{DC8F01}& $\times$ $\circ$ \cellcolor[HTML]{FF6B00}&\\
        (0,$P^{1*}_{FD}$,0,$P^{2*}_{FD}$) & \cellcolor[HTML]{64A8CC} & \cellcolor[HTML]{64A8CC} &  \cellcolor[HTML]{64A8CC}  & \cellcolor[HTML]{64A8CC} & \cellcolor[HTML]{64A8CC} & \cellcolor[HTML]{64A8CC} & \cellcolor[HTML]{64A8CC} & $\circ$ \cellcolor[HTML]{DC8F01}& $\times$ \cellcolor[HTML]{E9CA76}& $\times$ $\circ$ \cellcolor[HTML]{FF6B00}& $\times$  \cellcolor[HTML]{E9CA76} & $\times$ $\circ$ \cellcolor[HTML]{FF6B00}&\\
        (1,$P^{1*}_{FD}$,$P^{2*}_{FC}$,1) & \cellcolor[HTML]{64A8CC} & \cellcolor[HTML]{64A8CC} & \cellcolor[HTML]{64A8CC} & \cellcolor[HTML]{64A8CC} & \cellcolor[HTML]{64A8CC} & \cellcolor[HTML]{64A8CC} & $\circ$ \cellcolor[HTML]{DC8F01}& \cellcolor[HTML]{64A8CC} & \cellcolor[HTML]{64A8CC} & \cellcolor[HTML]{64A8CC} & \cellcolor[HTML]{64A8CC} & \cellcolor[HTML]{64A8CC} & \\
        (0,$P^{1*}_{FD}$,$P^{2*}_{FC}$,0) & \cellcolor[HTML]{64A8CC} & \cellcolor[HTML]{64A8CC} & \cellcolor[HTML]{64A8CC} & \cellcolor[HTML]{64A8CC} & \cellcolor[HTML]{64A8CC} & \cellcolor[HTML]{64A8CC} & \cellcolor[HTML]{64A8CC} & \cellcolor[HTML]{64A8CC} & \cellcolor[HTML]{64A8CC} & \cellcolor[HTML]{64A8CC} & \cellcolor[HTML]{64A8CC} &\cellcolor[HTML]{64A8CC} & \\
        ($P^{1*}_{FC}$,1,1,$P^{2*}_{FD}$) & \cellcolor[HTML]{64A8CC} & \cellcolor[HTML]{64A8CC} & \cellcolor[HTML]{64A8CC} & \cellcolor[HTML]{64A8CC} & \cellcolor[HTML]{64A8CC} & \cellcolor[HTML]{64A8CC} & \cellcolor[HTML]{64A8CC} & \cellcolor[HTML]{64A8CC} & \cellcolor[HTML]{64A8CC} & \cellcolor[HTML]{64A8CC} & \cellcolor[HTML]{64A8CC} & \cellcolor[HTML]{64A8CC}&  \\
        ($P^{1*}_{FC}$,0,0,$P^{2*}_{FD}$) & \cellcolor[HTML]{64A8CC} & \cellcolor[HTML]{64A8CC} & \cellcolor[HTML]{64A8CC} & \cellcolor[HTML]{64A8CC} & \cellcolor[HTML]{64A8CC} & \cellcolor[HTML]{64A8CC} & $\circ$ \cellcolor[HTML]{DC8F01}& \cellcolor[HTML]{64A8CC} & \cellcolor[HTML]{64A8CC} & \cellcolor[HTML]{64A8CC}&  \cellcolor[HTML]{64A8CC}& \cellcolor[HTML]{64A8CC} &  \\
        (1,$P^{1*}_{FD}$,0,$P^{2*}_{FD}$) & \cellcolor[HTML]{64A8CC} & $\circ$ \cellcolor[HTML]{DC8F01}& \cellcolor[HTML]{64A8CC} & $\times$ $\circ$ \cellcolor[HTML]{FF6B00}& \cellcolor[HTML]{64A8CC} & $\circ$ \cellcolor[HTML]{DC8F01}& \cellcolor[HTML]{64A8CC} & \cellcolor[HTML]{64A8CC} & \cellcolor[HTML]{64A8CC} &  \cellcolor[HTML]{64A8CC}& \cellcolor[HTML]{64A8CC} &\cellcolor[HTML]{64A8CC} & \\
        ($P^{1*}_{FC}$,0,$P^{2*}_{FC}$,1) & \cellcolor[HTML]{64A8CC} & \cellcolor[HTML]{64A8CC} & $\circ$ \cellcolor[HTML]{DC8F01}& \cellcolor[HTML]{64A8CC} & $\times$ $\circ$ \cellcolor[HTML]{FF6B00}& $\circ$ \cellcolor[HTML]{DC8F01}& \cellcolor[HTML]{64A8CC} & \cellcolor[HTML]{64A8CC} & \cellcolor[HTML]{64A8CC} & \cellcolor[HTML]{64A8CC} & \cellcolor[HTML]{64A8CC} &\cellcolor[HTML]{64A8CC} & \\
        (0,$P^{1*}_{FD}$,1,$P^{2*}_{FD}$) & \cellcolor[HTML]{64A8CC} & \cellcolor[HTML]{64A8CC} & \cellcolor[HTML]{64A8CC} & $\times$ $\circ$ \cellcolor[HTML]{FF6B00}& \cellcolor[HTML]{64A8CC} & \cellcolor[HTML]{64A8CC} & \cellcolor[HTML]{64A8CC} & \cellcolor[HTML]{64A8CC} & \cellcolor[HTML]{64A8CC} & \cellcolor[HTML]{64A8CC} & \cellcolor[HTML]{64A8CC} & \cellcolor[HTML]{64A8CC}& \\
        ($P^{1*}_{FC}$,1,$P^{2*}_{FC}$,0) & \cellcolor[HTML]{64A8CC} & \cellcolor[HTML]{64A8CC} & \cellcolor[HTML]{64A8CC} & \cellcolor[HTML]{64A8CC} & $\times$ $\circ$ \cellcolor[HTML]{FF6B00} & \cellcolor[HTML]{64A8CC} & \cellcolor[HTML]{64A8CC} & \cellcolor[HTML]{64A8CC} & \cellcolor[HTML]{64A8CC} & \cellcolor[HTML]{64A8CC} & \cellcolor[HTML]{64A8CC} &\cellcolor[HTML]{64A8CC} & \\
        ($P^{1*}_{FC}$,0,1,$P^{2*}_{FD}$) & $\times$ $\circ$ \cellcolor[HTML]{FF6B00} & $\circ$ \cellcolor[HTML]{DC8F01} & \cellcolor[HTML]{64A8CC}& \cellcolor[HTML]{64A8CC} & \cellcolor[HTML]{64A8CC} & \cellcolor[HTML]{64A8CC} & \cellcolor[HTML]{64A8CC} & \cellcolor[HTML]{64A8CC} & \cellcolor[HTML]{64A8CC} & \cellcolor[HTML]{64A8CC} & \cellcolor[HTML]{64A8CC} &\cellcolor[HTML]{64A8CC} & \\
        ($P^{1*}_{FC}$,1,0,$P^{2*}_{FD}$) & $\times$ $\circ$ \cellcolor[HTML]{FF6B00} & $\times$ \cellcolor[HTML]{E9CA76}& $\times$ \cellcolor[HTML]{E9CA76}& \cellcolor[HTML]{64A8CC} & $\circ$ \cellcolor[HTML]{DC8F01} & $\times$ \cellcolor[HTML]{E9CA76}& \cellcolor[HTML]{64A8CC} & \cellcolor[HTML]{64A8CC} & \cellcolor[HTML]{64A8CC} & \cellcolor[HTML]{64A8CC} & \cellcolor[HTML]{64A8CC} &\cellcolor[HTML]{64A8CC} &\\
        (1,$P^{1*}_{FD}$,$P^{2*}_{FC}$,0) & $\times$ $\circ$ \cellcolor[HTML]{FF6B00}  & \cellcolor[HTML]{64A8CC} & \cellcolor[HTML]{64A8CC} & $\circ$ \cellcolor[HTML]{DC8F01} & \cellcolor[HTML]{64A8CC} & \cellcolor[HTML]{64A8CC} & \cellcolor[HTML]{64A8CC} & \cellcolor[HTML]{64A8CC} & \cellcolor[HTML]{64A8CC} & \cellcolor[HTML]{64A8CC} & \cellcolor[HTML]{64A8CC} &\cellcolor[HTML]{64A8CC} & \\
        (0,$P^{1*}_{FD}$,$P^{2*}_{FC}$,1) & $\times$ $\circ$ \cellcolor[HTML]{FF6B00} & $\times$ $\circ$ \cellcolor[HTML]{FF6B00} & $\times$ \cellcolor[HTML]{E9CA76} & \cellcolor[HTML]{64A8CC} & \cellcolor[HTML]{64A8CC} & $\times$ \cellcolor[HTML]{E9CA76}& \cellcolor[HTML]{64A8CC} & \cellcolor[HTML]{64A8CC} & \cellcolor[HTML]{64A8CC} & \cellcolor[HTML]{64A8CC} & \cellcolor[HTML]{64A8CC} &\cellcolor[HTML]{64A8CC} &
    \end{tabular}
    \caption{Regions of Fig. \ref{fig:PC2_combined} where a given response strategy is an equilibrium for the PC game with C preference ($\times$) and the BoS game ($\circ$) .} 
    \label{tab:PC2_combined}
    \label{equilibriumtabel}
\end{table*}

In Table \ref{equilibriumtabel} the equilibria that hold in each region of Fig. \ref{fig:PC2_combined} are further specified. We divide the equilibria in three classes: class $1$ represents the strategies that are equilibria in all regions; class $2$ indicates the remaining strategies that consist only of pure strategies; class $3$ distinguishes the strategies that contain both pure and mixed strategies.

\subsubsection{Payoff Analysis}

After we establish which response strategies are equilibria in each region representing the initial correlation devices, we calculate which ones give the highest payoffs, as the players are interested in obtaining the highest outcome on a stable configuration. For the players of the PC game, the rational choice is to ignore the correlations and coordinate on their preference. On the contrary, players on a BoS game stand to benefit from the correlations, because they want to prevent as much as possible to carry out an action alone, be it their preference or that of their opponents. Considering that the players are not symmetric, the best outcomes can happen when distinct equilibria are adopted by different players in regions where multiple equilibrium solutions exist. Considering the class $1$ equilibria, we find that the pure C equilibrium $(1,0,1,0)$ is best for player $1$, whereas the pure D equilibrium $(0,1,0,1)$ is best for player $2$, with the payoff of the most disadvantaged player equal to $s$. These give the best possible expected payoffs and are equilibria in all regions, but the issues of miscoordination remain. The mixed strategy equilibrium is not parameterized by symmetric probabilities, since $p_{CD}=\left(1/(s+1)\right)^2$ and $p_{DC}=\left(s/(s+1)\right)^2$, but it is always achievable if the players choose to ignore the correlations. However, it has the lowest expected payoff compared to any other equilibrium solution. 

Wherever the response strategies of class $2$ are equilibria in the same regions as equilibria of class $3$, the former always present a higher payoff. This happens in regions A, B, C and L for the BoS game, the regions highlighted in Fig. \ref{fig:maxpayoff}. In region $1$ of this figure, there are two degenerate equilibria, equally profitable for both players. This happens in spite of the fact that the renormalization schemes that apply to this region do not preserve the symmetry of the initial probabilities. As an example, consider the symmetric initial correlation device $p_{CC}=1/5$, $p_{CD}=p_{DC}=2/5$ and $p_{DD}=0$. Applying the response strategy $(1,1,0,0)$ will renormalize the correlation device to $p^R_{CC}=p^R_{DD}=2/5$, $p_{CD}=1/5$ and $p_{DC}=0$, whereas the response strategy $(0,0,1,1)$ will transform the initial correlation device to $p^R_{CC}=p^R_{DD}=2/5$, $p_{CD}=0$ and $p_{DC}=1/5$. However, in spite of the different renormalized correlations, the payoff of the BoS game only has non-zero values when the players coordinate. Therefore, the expected payoff after applying these response strategies will be identical for both players. 

In regions $2$ and $3$, only one of these two strategies is an equilibrium, and so the players agree on the strategy to adopt and the benefits of introducing the correlations are clear. In regions $4$ and $5$, the two players have different payoffs for different strategies, since strategies $(1,1,1,1)$ and $(0,0,0,0)$ reverse the diagonal probability values, so although both are equilibria, there exists an undecidability equivalent to reaching a pure solution without correlations. 

\begin{figure*}[t]
\centering
\begin{subfigure}{.3\textwidth}
 \centering
 \includegraphics[width=.9\linewidth]{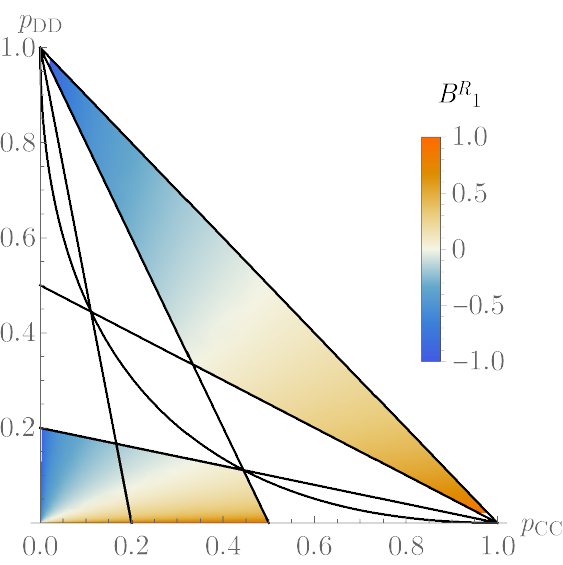}
 \caption{}
 \label{fig5.1b}
 \end{subfigure}
 \begin{subfigure}{.3\textwidth}
  \centering
  \includegraphics[width=.9\linewidth]{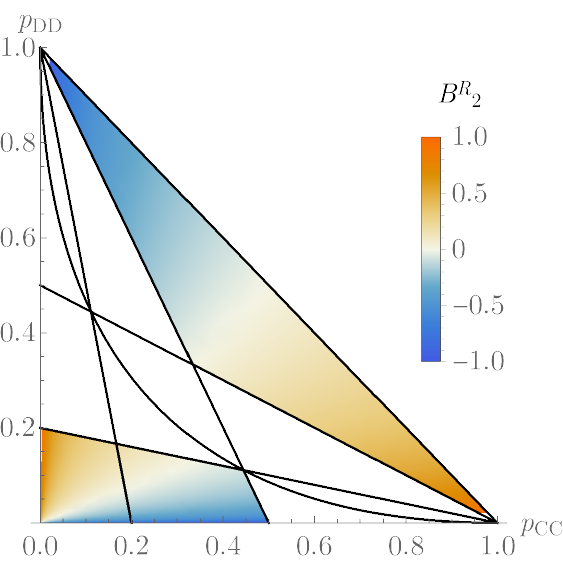}
  \caption{}
  \label{fig5.1b2}
\end{subfigure}

    \begin{subfigure}{.3\textwidth}
  \centering
  \includegraphics[width=.9\linewidth]{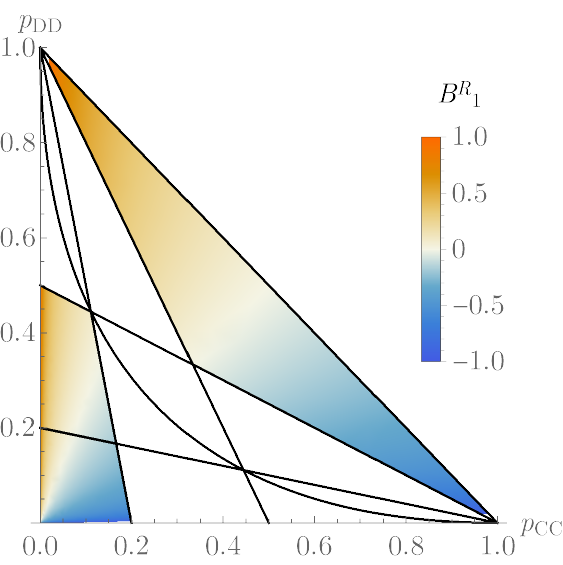}
  \caption{}
  \label{fig5.1a}
\end{subfigure}%
\begin{subfigure}{.3\textwidth}
  \centering
  \includegraphics[width=.9\linewidth]{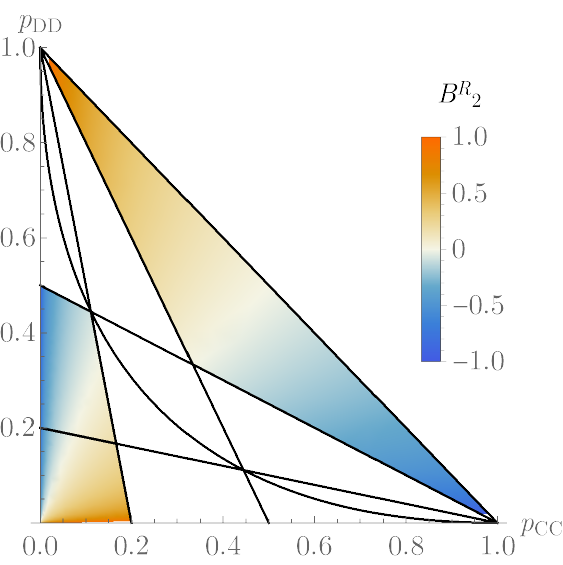}
  \caption{}
  \label{fig5.1a2}
\end{subfigure}%
\begin{subfigure}{.3\textwidth}
  \centering
  \includegraphics[width=.9\linewidth]{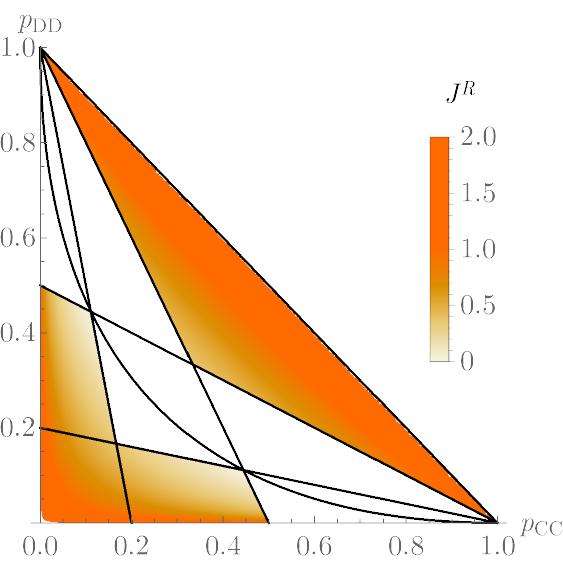}
  \caption{}
  \label{fig5.1b}
\end{subfigure}
\caption{Renormalized Ising parameters using the class $2$  equilibria for the BoS game as a function of symmetric initial correlations, at $s=1/2$. \textbf{a) and b):} Renormalized magnetic fields, using the (1,1,1,1) and (1,1,0,0) equilibrium strategies, for a) player $1$ and b) player 2. \textbf{c) and d):} Renormalized magnetic fields, using the (0,0,0,0) and (0,0,1,1) equilibrium strategies, for c) player $1$ and d) player 2. \textbf{e):} Renormalized interaction strength.}
\label{fig:5.1}
\end{figure*}

\subsection{Ising Model}

\begin{figure*}[t]
\centering
\begin{subfigure}{.3\textwidth}
 \centering
 \includegraphics[width=.9\linewidth]{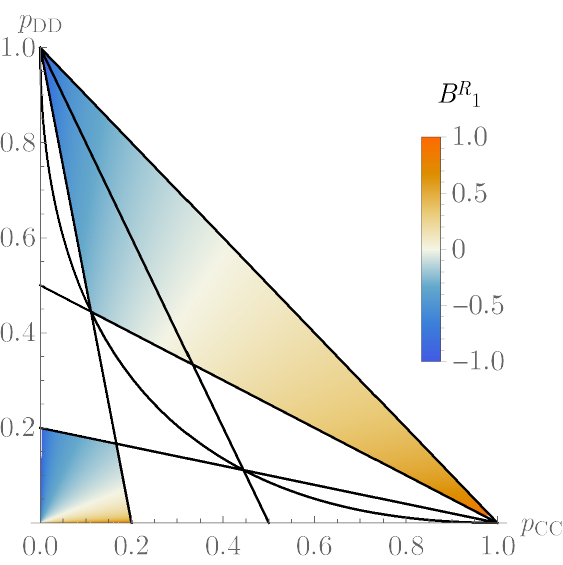}
 \caption{}
 \label{fig5.1b}
 \end{subfigure}
 \begin{subfigure}{.3\textwidth}
  \centering
  \includegraphics[width=.9\linewidth]{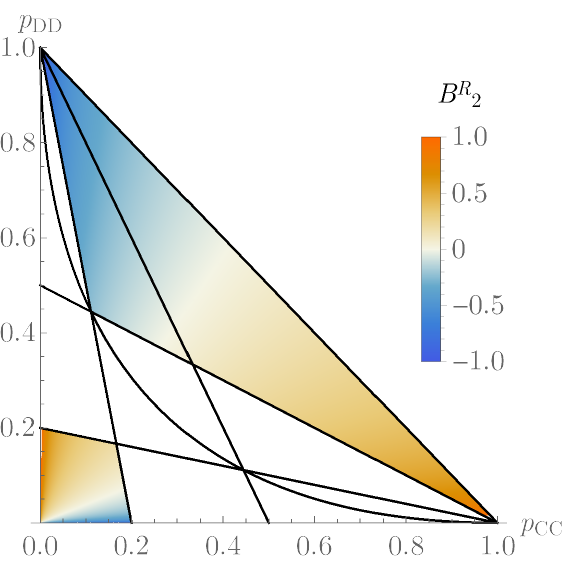}
  \caption{}
  \label{fig5.1b2}
\end{subfigure}

\begin{subfigure}{.3\textwidth}
  \centering
  \includegraphics[width=.9\linewidth]{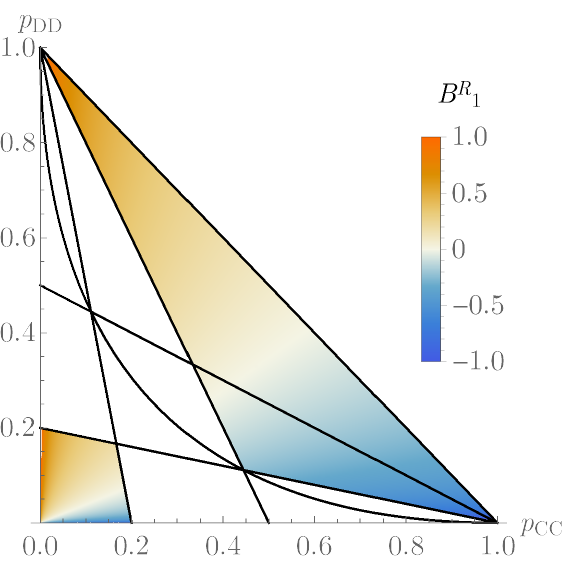}
  \caption{}
  \label{fig5.1a}
\end{subfigure}%
\begin{subfigure}{.3\textwidth}
  \centering
  \includegraphics[width=.9\linewidth]{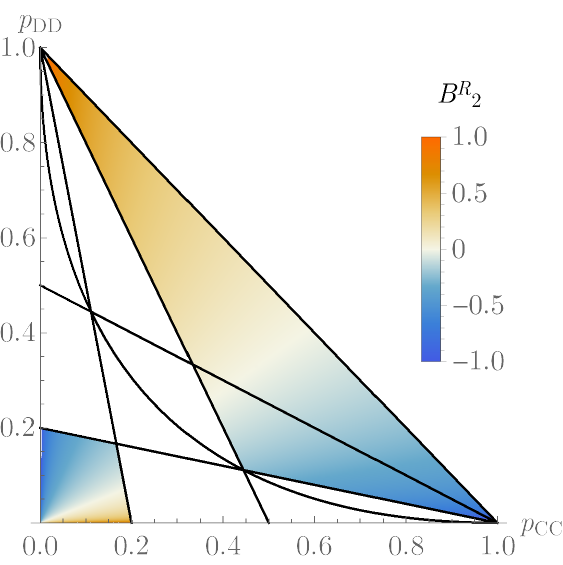}
  \caption{}
  \label{fig5.1a2}
\end{subfigure}%
\begin{subfigure}{.3\textwidth}
  \centering
  \includegraphics[width=.9\linewidth]{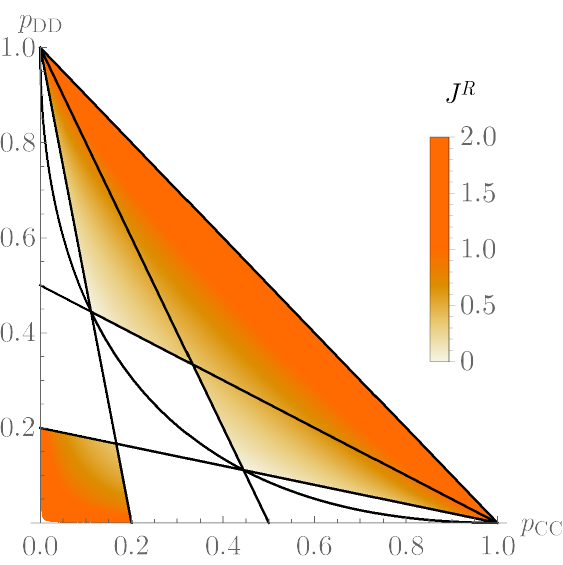}
  \caption{}
  \label{fig5.1b}
\end{subfigure}
\caption{Renormalized Ising parameters using the class $2$  equilibria for the PC game with C preference as a function of symmetric initial correlations, at $s=1/2$. \textbf{a) and b):} Renormalized magnetic fields, using the (1,1,1,1) and (1,1,0,0) equilibrium strategies, for a) player $1$ and b) player 2. \textbf{c) and d):} Renormalized magnetic fields, using the (0,0,0,0) and (0,0,1,1) equilibrium strategies, for c) player $1$ and d) player 2. \textbf{e):} Renormalized interaction strength.}
\label{fig:5.12}
\end{figure*}

Going into correlated games, we translate the results obtained using the slope analysis for two players and describe them as functions of the Ising parameters. Both initial and renormalized probabilities are respectively mapped to the Ising Hamiltonians $H_{\mu_1\mu_2}$ and $H^R_{\mu_1\mu_2}$. 
Inverting the expressions in Eq.(\ref{initialproabilities}), we obtain the Ising parameters as functions of the initial probabilities as

\begin{align} \label{ising_mapping2D}
&B_1 = \frac{1}{4}\log\left(\frac{p_{CC}p_{CD}}{p_{DD}p_{DC}}\right), \nonumber \\
&B_2 = \frac{1}{4}\log\left(\frac{p_{CC}p_{DC}}{p_{DD}p_{CD}}\right),   \nonumber \\
&J = \frac{1}{4}\log\left(\frac{p_{CC}p_{DD}}{p_{CD}p_{DC}}\right).
\end{align} The renormalized probabilities are mapped similarly to the parameters $B^R_1$, $B^R_2$ and $J^R$.

From Eqs.(\ref{ising_mapping2D}), we see that in the symmetric section of the BoS game, shown in Fig. \ref{symmetricfig}, the relationship between the magnetic fields is such that $B_1=B_2$. This means that, from all uncorrelated strategies, only the pure strategies will be represented in this section, since the mixed Nash equilibria of BoS is obtained with parameters $J=0$, $B_1=-\ln(s)/2$ and $B_2=-B_1$. The null value of the interaction strength confirms the independence of each of the players, and the anti-symmetric values of magnetic field strength mirror their different preferences. These are obtained as the modulo of the magnetic field becomes large, with a positive sign originating the result where both play C ($p_{CC}=1$) and a negative sign where they play D ($p_{DD}=1)$.

After we perform the slope analysis, we are interested in understanding how the Ising parameters change as a function of their initial values, that describe the initial probabilities, and of the renormalization strategies that are available for each game in the several regions. To that effect, we plot the parameters $B_1^R$, $B_2^R$ and $J^R$ in Figs. \ref{fig:5.1} and \ref{fig:5.12} for the BoS and the PC games, respectively. This graphically characterizes the renormalized correlations given by the class $2$ strategies that are equilibria in each region, as functions of symmetric initial correlations. As explained above, response strategies $(1,1,0,0)$ and $(0,0,1,1)$ prompt a set of asymmetric renormalized probabilities, reflected in the different magnetic fields assigned to each player in the regions where these strategies are equilibria. Notably, $J^R$ is always positive in the entire plane, which is an indication of the preference for ferromagnetic alignment in all equilibria, as is desirable for the coordination games studied here.

\section{Response Strategies for Three-Player Games}\label{threeplayergames}

\begin{figure*}[t]
\centering
\begin{subfigure}{.3\textwidth}
  \centering
  \includegraphics[width=.8\linewidth]{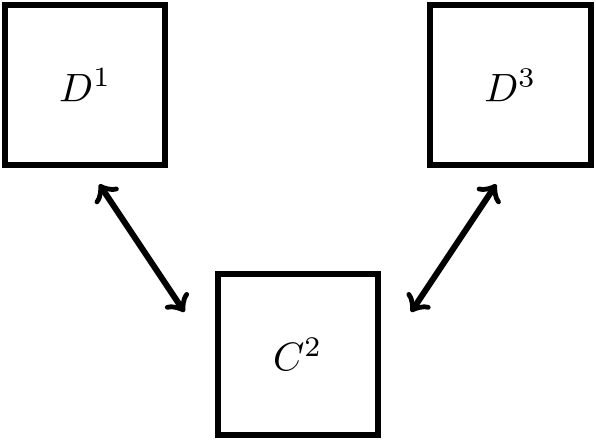}
  \caption{D-C-D}
  \label{fig4.1h}
\end{subfigure}%
\begin{subfigure}{.3\textwidth}
 \centering
  \includegraphics[width=.8\linewidth]{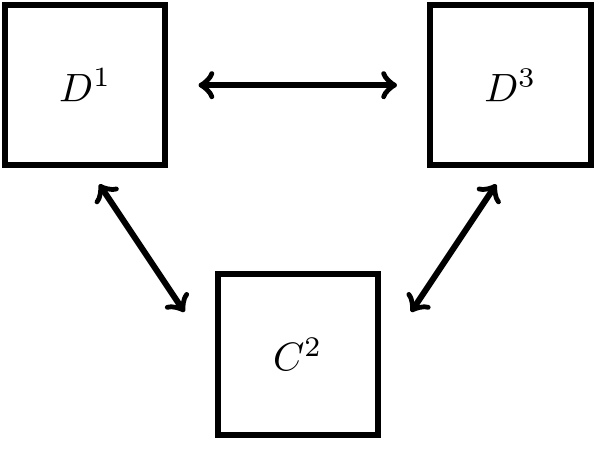}
  \caption{-D-C-D-}
  \label{fig4.1e}
\end{subfigure}%
\begin{subfigure}{.3\textwidth}
 \centering
  \includegraphics[width=.8\linewidth]{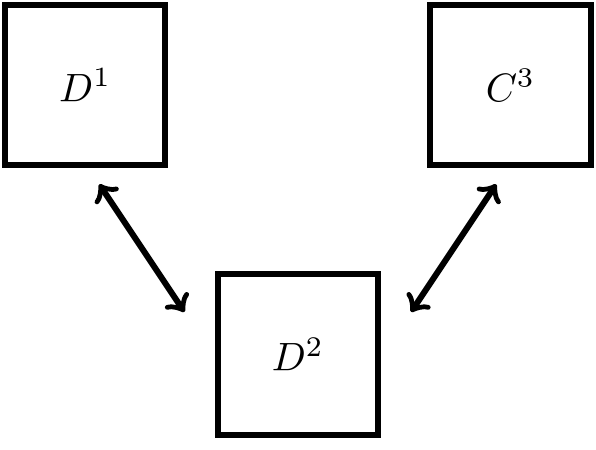}
  \caption{D-D-C}
  \label{fig4.1aa}
\end{subfigure}%
\caption{Different three-player networks with two players with a D preference and one player with C preference.}
\label{3pnetworks}
\end{figure*}

In this section we analyze the Nash equilibria corresponding to the three networks, shown in Figs. \ref{3pnetworks}, that form substructures of the networks in Figs. \ref{Networks}. After describing how the correlations from the previous section extend to three players, we map them to a generalized Ising model and impose simplifying assumptions over the Ising variables, as the full phase space of the three-player games is too large to explore here in full detail. We explore how these initial variables change under several renormalization schemes. Finally, we compare the correlated equilibrium regions for each network as functions of the Ising parameters.

\subsection{Correlations on networks}

Now that we have found the different Hamiltonians that describe the equilibrium states for the two-player games, we might be tempted to extent them to every single interaction in a larger network, inferring that they are governed by the same Hamiltonians. However, unless all the nodes in the network have the same preference, the conflicting interactions that the nodes have with their neighbors will originate new global equilibrium states, as we will see. For this reason, we will need to introduce a generalization to the Ising model, accounting for long-range effects.

We start by assuming that a set of global three-player configurations exists. These must depend on the number and type of local correlation, and therefore be emergent from the correlation devices modeling the update rules. Local correlations can be parametrized by the same Ising parameters of a global correlation. This is especially useful to gain insight about how any two players are correlated, beyond those that are direct opponents. These correlations do not contribute to the establishment of an equilibrium of the network, since these players do not look at each other to calculate their payoff and update their choices. They are, instead, a byproduct of the games that other players play against each other locally, but they are covered by the global correlation device. Therefore, the Ising model offers an advantage in studying the equilibria, since it only \textit{describes} statistically any set of joint outcomes, but we still need the correlation device in first instance to calculate the Nash equilibria using the slope analysis, because it \textit{prescribes} the possible outcomes of the players.

\subsubsection{Local Correlations}

Starting from a general three-player probability distribution 

\begin{equation}\label{3pprobs}
    p^R_{\mu_1\mu_2\mu_3}=\sum_{\mu_1'\mu_2'\mu_3'}P^1_{\mu_1\leftarrow\mu'_1}P^2_{\mu_2\leftarrow\mu_2'}P^3_{\mu_3\leftarrow\mu'_3}p_{\mu'_1\mu'_2\mu'_3},
\end{equation} we sum over the outcomes of one of the players to find the correlations between the other two. For instance, the effective correlations for players $1$ and $2$ coming from Eq.(\ref{3pprobs}) are given by

\begin{align}
   & p^{R}_{\mu_1,\mu_2} = \sum_{\mu_3} p^R_{\mu_1\mu_2\mu_3} = \sum_{\mu'_1\mu_2'}P^1_{\mu_1\leftarrow\mu_1'}P^2_{\mu_2\leftarrow\mu_2'}p_{\mu'_1\mu'_2}.
\end{align} Similarly, we obtain $p^{R}_{\mu_1,\mu_3}$ for players $1$ and $3$  and $p^{R}_{\mu_1,\mu_3}$ for players $2$ and $3$ by summing over $\mu_2$ and $\mu_1$, respectively. We note that first renormalizing and then tracing out the actions of one of the players is equivalent to the procedure in reversed order. In this way, we can apply the slope analysis to renormalize the correlations, using the two-player correlations as correlation devices if there is a link. Thus, we extend the slope analysis to the three-player case using  Eq.(\ref{slopepayoff}), assuming that the players respond by following C or D with respect to the instructions of the three-player correlation device.

The response strategies now become the tuples $(P^1_{FC}, P^1_{FD}, P^2_{FC}, P^2_{FD}, P^3_{FC}, P^3_{FD})$. A response strategy is an equilibrium if and only if all six corresponding slope conditions are met simultaneously. These conditions assume that the incomplete network knowledge possessed by one player consists of her own preference and her degree of connectedness. This ensures that conditions are specific to every network structure. 

\subsubsection{Generalized Ising model}

We map the three-player correlation probabilities to an extended Ising model. Dealing with the interactions of at least three particles, we can extend the Ising Hamiltonian by introducing a three-body interaction term $A$, resulting in
\begin{align}\label{hamiltonian3}
    H_{\mu_1 \mu_{2} \mu_{3}}= & -A \mu_1 \mu_{2} \mu_{3} \nonumber \\
    &-  J_{12}\mu_1 \mu_{2} - J_{23}\mu_2 \mu_{3} - J_{13}\mu_1 \mu_{3} \nonumber \\
    &-  \sum_i B_i \mu_i
    .
\end{align} 
Expanding our model with the three-body term $A$ enables a unique translation between the seven degrees of freedom of the three-player correlation device (a probability distribution for each of the eight possible configurations minus the completeness relation) to the seven Ising parameters. In practice, the three-body interaction introduces true correlations between all players, which are then not product-separable, neither with individual nor with pair-wise terms.

\begin{table*}[t]\label{renormalizationparameters}
\centering
\renewcommand{\arraystretch}{1.5}
\setlength{\tabcolsep}{3pt}
\begin{tabular}{c|ccccccccc}
$\frac{J'}{J}$                  & \multicolumn{3}{c|}{$-1$}                        & \multicolumn{3}{c}{$0$}                                                     & \multicolumn{3}{c}{$1$}                                \\ \hline
$A$                  & $<0 $   & $=0$    & \multicolumn{1}{c|}{$>0$}    & $<0 $            & $=0$             & \multicolumn{1}{c|}{$>0$}             & $<0 $            & $=0$             & $>0$             \\ \hline
$(1, 1, 1, 1, 1, 1)$ & $\circ$ \cellcolor[HTML]{DC8F01}& $\circ$ \cellcolor[HTML]{DC8F01} & \multicolumn{1}{c|}{$\circ$ \cellcolor[HTML]{DC8F01}} & \cellcolor[HTML]{FF6B00} $\times$ $\circ$ $\square$  \cellcolor[HTML]{FF6B00}& $\times$ $\circ$ $\square$  \cellcolor[HTML]{FF6B00}& \multicolumn{1}{c|}{$\times$ $\circ$ $\square$  \cellcolor[HTML]{FF6B00}} & $\times$ $\circ$ $\square$  \cellcolor[HTML]{FF6B00} & $\times$ $\circ$ $\square$  \cellcolor[HTML]{FF6B00} & $\times$ $\circ$ $\square$  \cellcolor[HTML]{FF6B00}\\
$(0, 0, 0, 0, 0, 0)$ & $\circ$ \cellcolor[HTML]{DC8F01}& $\circ$ \cellcolor[HTML]{DC8F01}& \multicolumn{1}{c|}{$\circ$ \cellcolor[HTML]{DC8F01}} & $\times$ $\circ$ $\square$  \cellcolor[HTML]{FF6B00}& $\times$ $\circ$ $\square$  \cellcolor[HTML]{FF6B00}& \multicolumn{1}{c|}{$\times$ $\circ$ $\square$  \cellcolor[HTML]{FF6B00}} & $\times$ $\circ$ $\square$  \cellcolor[HTML]{FF6B00} & $\times$ $\circ$ $\square$  \cellcolor[HTML]{FF6B00}& $\times$ $\circ$ $\square$  \cellcolor[HTML]{FF6B00}\\
$(1, 1, 0, 0, 1, 1)$ & $\times$  $\circ$ $\square$  \cellcolor[HTML]{FF6B00}& $\times$  $\circ$ $\square$  \cellcolor[HTML]{FF6B00}& \multicolumn{1}{c|}{$\times$  $\circ$ $\square$  \cellcolor[HTML]{FF6B00}} &$\times$  $\circ$ $\square$  \cellcolor[HTML]{FF6B00}         &$\times$  $\circ$ $\square$  \cellcolor[HTML]{FF6B00}        & \multicolumn{1}{c|}{$\times$ $\circ$ $\square$  \cellcolor[HTML]{FF6B00}}          &     \cellcolor[HTML]{64A8CC}             &   \cellcolor[HTML]{64A8CC}               &  $\circ$ \cellcolor[HTML]{DC8F01}\\
$(0, 0, 1, 1, 0, 0)$ & $\times$  $\circ$ $\square$  \cellcolor[HTML]{FF6B00}& $\times$  $\circ$ $\square$ \cellcolor[HTML]{FF6B00}& \multicolumn{1}{c|}{$\times$ $\circ$ $\square$  \cellcolor[HTML]{FF6B00}} &$\times$  $\circ$ $\square$  \cellcolor[HTML]{FF6B00}         & $\times$ $\circ$ $\square$  \cellcolor[HTML]{FF6B00}         & \multicolumn{1}{c|}{$\times$ $\circ$ $\square$  \cellcolor[HTML]{FF6B00}}          &  $\circ$ \cellcolor[HTML]{DC8F01} &      \cellcolor[HTML]{64A8CC}            &     \cellcolor[HTML]{64A8CC}            
\end{tabular}
\caption{Parameters for which a region in the $B-J$ plane, with $B,J \in [-10,10]$, exists, such that a response strategy is an equilibrium strategy. The networks are the D-C-D network in Fig. \ref{fig4.1h} ($\circ$), the -D-C-D- network in Fig. \ref{fig4.1e} ($\times$) and the D-D-C network in Fig. \ref{fig4.1aa}, for $s=1/2$ and $B'/B=-1$. The last two strategies are equivalent to $(1,1,1,1,0,0)$ and $(0,0,0,0,1,1)$ in the D-D-C network, respectively.}
\label{equilibriumexistance}
\end{table*}

There exist $10$ different three-player networks, depending on the existence of two or three links between nodes and counting with two preferences per node. We study here only the equilibria on networks where one of the players has a different preference with respect to the other two, as schematized in Fig. \ref{3pnetworks}. Moreover, for simplicity we use the network structures to impose relationship between the bare parameters. We define the same interaction $J$ between players with different preferences, and an interaction $J'$ between players with the same preference. Furthermore, we define a magnetic field $B$ for players that prefer C, and $B'$ for those preferring D. For the networks in Figs. \ref{fig4.1h} and \ref{fig4.1e}, we thus assign $J_{12}=J_{23}=J$, $J_{13}=J'$, $B_1=B_3=B'$ and $B_2=B$, while, for the network in Fig. \ref{fig4.1aa}, $J_{23}=J_{13}=J$, $J_{12}=J'$, $B_1=B_2=B'$ and $B_3=B$. 

Using this parametrization for the initial correlation device, a comparison between networks D-C-D and -D-C-D- is particularly relevant, as it tells us how strongly it depends on the existence or absence of a game played between players $1$ and $3$. An indirect correlation between the outcomes of these players can emerge, dependent on their direct correlation with player $2$, and so described by the ratio $J'/J$. Due to the preferences of these players, this parametrization furthermore reflects the relative interaction strength between the BoS and the PC game in case there is a link between them.

We look closely at the cases where different relative preferences induce one ferromagnetic and one antiferromagnetic alignment ($J'/J=-1$), the cases where all interactions are either ferromagnetic or anti-ferromagnetic ($J'/J=1$) and, finally, also the cases where two of the players are always uncorrelated ($J'/J=0$). A similar analysis takes place for the three-body interaction $A$, for which the particular values $A=-1/2$, $A=0$ and $A=1/2$ represent the characteristic effects of a sign change in this parameter. We also assume that the probabilities are not symmetric, in particular imposing that $B'/B=-1$, since the preference of a player is connected with the direction of the external magnetic field. This allows us also to recover the mixed strategy solution in this setting.

We focus the analysis of the three-player game on the extension of the two-player class $2$ equilibria, as these were found to be the best solutions for symmetric initial correlations in the BoS game. In this way, we can study their behaviour in a wider range of initial probabilities. For networks D-C-D and -D-C-D- these class $2$ Nash Equilibria will be $(1,1,1,1,1,1)$ (the correlated equilibrium), $(0,0,0,0,0,0)$, $(1,1,0,0,1,1)$ and $(0,0,1,1,0,0)$. Network D-D-C shares the first two equilibria, but the last two are replaced by $(1,1,1,1,0,0)$ and $(0,0,0,0,1,1)$. 
Table \ref{equilibriumexistance} shows a list of the class $2$ response strategies with equilibrium solutions in a certain range of parameters, for each of the networks in Fig. \ref{3pnetworks}. Network D-C-D is the one for which these strategies have the largest scope as solutions, because this network consists only of BoS games.

\subsection{Renormalization}

We find the existence of symmetries between the initial and renormalized Ising parameters as result of the application of the class $2$ response strategies, schematized in Table \ref{renormalizationparameters}.

\begin{table}
    \centering
\begin{tabular}{crrrrr}
                                          & \multicolumn{1}{c}{$J^R$} & \multicolumn{1}{c}{$J'^R/J^R$} & \multicolumn{1}{c}{$B^R$} & \multicolumn{1}{c}{$B'^R/B^R$} & \multicolumn{1}{c}{$A$} \\ \hline
\multicolumn{1}{c|}{$(1, 1, 1, 1, 1, 1)$} & $J$                     & $J'/J$                     & $B$                     & $B'/B$                     & $A$                     \\
\multicolumn{1}{c|}{$(0, 0, 0, 0, 0, 0)$}  & $J$                     & $J'/J$                     & $-B$                     & $B/B'$                     & $-A$                   \\
\multicolumn{1}{c|}{$(1, 1, 0, 0, 1, 1)$} &  $-J$                     & $-J'/J$                     & $B$                     & $-B'/B$                     & $-A$                    \\
\multicolumn{1}{c|}{$(0, 0, 1, 1, 0, 0)$} &  $-J$                     & $-J'/J$                     & $-B$                     & $-B'/B$                     & $A$                    
\end{tabular}
    \caption{Renormalization of Ising parameters given by the equilibrium response strategy. This is true for all studied networks except for network D-D-C, for which the last two parameter renormalizations are obtained with strategies  $(1,1,1,1,0,0)$ and $(0,0,0,0,1,1)$ instead.}
    \label{renormalization}
    \label{renormalizationparameters}
\end{table}

\begin{figure*}[t]
    \centering
    \begin{subfigure}{.3\textwidth}
  \centering
  \includegraphics[width=.9\linewidth]{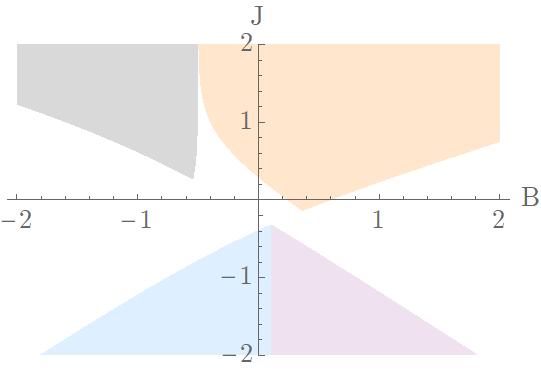}
  \caption{$A=-1/2$}
  \label{fig5.1a}
\end{subfigure}%
\begin{subfigure}{.3\textwidth}
  \centering
  \includegraphics[width=.9\linewidth]{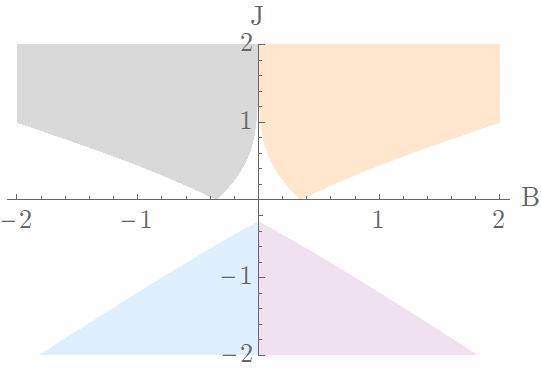}
  \caption{$A=0$}
  \label{fig5.1b}
\end{subfigure}
\begin{subfigure}{.3\textwidth}
  \centering
  \includegraphics[width=.9\linewidth]{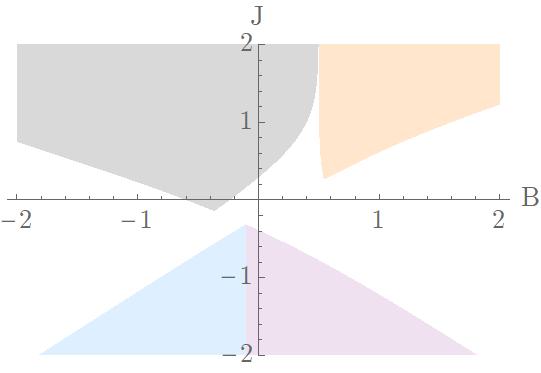}
  \caption{$A=1/2$}
  \label{fig5.1b}
\end{subfigure}
\caption{Regions in the $B-J$ plane where the class $2$ response strategies are equilibria of the game played on the D-C-D network, for $s=1/2$, $J'/J=-1$ and $B'/B=-1$. Orange represents strategy $(1, 1, 1, 1, 1, 1)$  and gray represents strategy $(0, 0, 0, 0, 0, 0)$. In purple and blue the two strategies $(1, 1, 0, 0, 1, 1)$ and $(0, 0, 1, 1, 0, 0)$ are equilibria, the different colours indicating different preferences for different players regarding payoff.}
\label{DCDf=-1}
\label{D-C-Df=-1}
\end{figure*} The knowledge of these relationships is useful, as we only need to apply the slope analysis to one of the response strategies and compute the initial correlations where it is in equilibrium, with the other three following suit.   
 As an example, in Fig. \ref{D-C-Df=-1} we zoom in on the equilibrium regions of network D-C-D, as a function of the initial probability Ising parameters, specifically with $J'/J=-1$ and $B'/B=-1$. As we go from $A$ to $-A$, we see the regions where $(1,1,1,1,1,1)$ and $(0,0,0,0,0,0)$ are Nash equilibria transform into one another upon a symmetry about the $J$ axis, and similarly between the strategies $(1,1,0,0,1,1)$ and $(0,0,1,1,0,0)$. Going from one of the latter strategies to one of the former cannot be represented in this plot, as it maps to $B/B'=1$, but we can already see the bulk of the equilibrium region going from $J$ to $-J$. With respect to payoff, we find a situation similar to region $1$ in Fig. \ref{fig:maxpayoff}, where there is more than one possible equilibrium solution. The players with preference for D will have a higher payoff adopting strategy $(1,1,0,0,1,1)$ on the blue region and strategy $(0,0,1,1,0,0)$ on the brown region, while player C will have the highest payoff choosing opposite strategies. In all cases, the payoff is higher than the mixed strategy solution.

\subsection{Comparing Networks}\label{comparingnetworks}

\begin{figure*}[t]
    \centering
\begin{subfigure}{\textwidth}
\centering
    \includegraphics{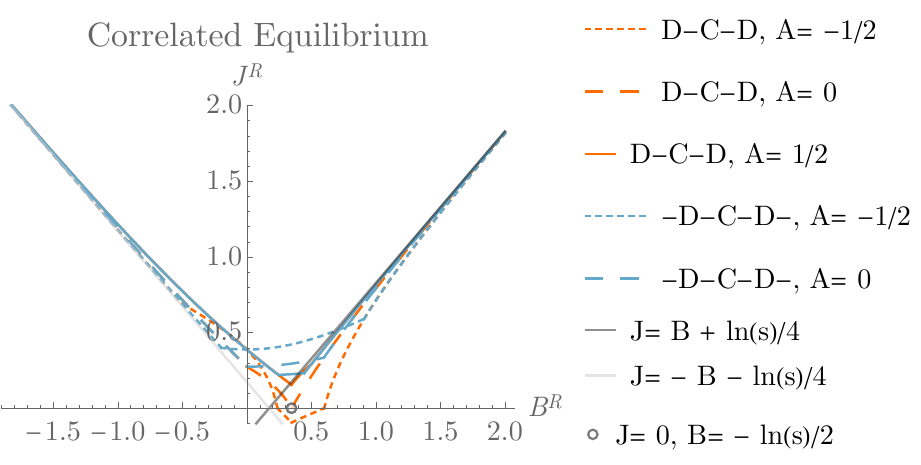}
    \subcaption{}
    \label{fig:correlatedf=1}
\end{subfigure}%

\begin{subfigure}{\textwidth}
  \centering
    \includegraphics{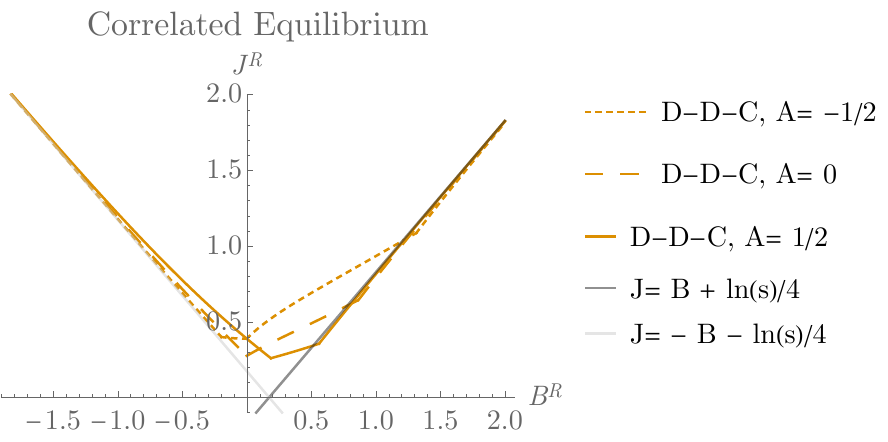}
    \subcaption{}
    \label{correlatedDDC}
\end{subfigure}%
\caption{Correlated equilibrium regions of networks D-C-D and -D-C-D- in (a) and of network D-D-C in (b), located above the full lines for $A <0$, the dashed lines for $A= 0$ and the dotted lines for $A >0$, for $s= 1/2$, $J'=J$ and $B'=-B$. In (a) the regions are bounded by the limiting behavior of the equilibrium conditions of player 2, namely for following C (dark gray) and for following D (light gray), and in (b) by the limiting behavior of the equilibrium conditions of players $1$ (dark gray) and $3$ (light gray) for following D.}
\end{figure*}

As there is a symmetry between the different renormalization schemes, we restrict our comparison between networks to the correlated equilibrium regions. The comparison between the correlated equilibrium regions of the D-C-D and -D-C-D- networks in Fig. \ref{fig:correlatedf=1} shows the difference between players $1$ and $3$ directly or indirectly playing against each other.

Here the choice of $J'/J=1$ reflects an incentive for alignment. For $J$ and $B$ large, we observe a converging behavior that is common to both networks and independent of the parameter $A$, and is fully determined by the equilibrium conditions of player $2$. To see this, consider the slope of the payoff function of player $2$ with respect to following the instruction to play $C$, towards correlated equilibrium. The corresponding condition $C^2_C > 0$ on networks -D-C-D- and D-C-D can be written, for $J'=J$ and $B'=-B$, as

\begin{equation}
 J > \frac{1}{4} \left[B-A + \ln \left( e^{B-A} (s-1) + e^{3B+A} s \right)\right ]. 
\end{equation} In the limit of large B, this condition can be reduced to the asymptote by first using that $B \gg A$, an then by keeping only the largest term in $B$. Ultimately this gives

\begin{align}
 J > & \frac{1}{4} \left[B+ \ln \left( e^{B} (s-1) + e^{3B} s \right)\right ] \nonumber  \\
   & \approx  B + \frac{\ln(s)}{4}. 
\end{align} In a similar way, the limiting behaviour around $J>-B-\ln(s)/4$ can be obtained from $C^2_D > 0$. This asymptotic behavior is shared by network D-D-C with the previous two networks. However, differentely from them, it is related with the equilibrium conditions of the peripheral nodes, obtained from $C^1_D>0$ and $C^3_D>0$, respectively. As with the fully correlated network, the renormalized interaction strength for this network is always positive. 

In contrast, for small values of $J$ and $B$, the value of $A$ becomes increasingly relevant. The limits of the stability region are given by the equilibrium conditions of players $1$ and $3$, where the dependence of the equilibria on the linking becomes apparent. The symmetry is higher for the closed network. 

We can also locate the standard Nash equilbria (the pure strategy C, where $p_{CCC}=1$; the pure strategy D, where $p_{DDD}=1$; and the mixed strategy) on these regions. We find the first pure strategy above the asymptote $J=-B- \ln(s)/4$ for large $-B$ and the second pure strategy at $J=B+ \ln(s)/4$ for large $B$, for all networks. The dependence of $J$ on $B$ relates with $B'=-B$. Given our parametrization, a large $|B|$ forces two particles into one spin direction but a third one into another, inducing opposite alignments. To counter this effect, $J$ has to be sufficiently large. The restriction of $\ln(s)/4$ emerges directly from the computation of the correlation probabilities using the slope analysis, thus directly encoding information about the payoff structure.

The mixed strategy is only found for network D-C-D and when $A=0$, for the values $J=0$ and $B=\ln(s)/2$, consistent with this being the equilibrium solution of a fully uncorrelated BoS game. It is found when both the two- and three-player interaction parameters are zero and $B'/B=-1$, with the $B$ parameter assuming the same value as in the mixed strategy of the BoS game. The mixed strategy is not present in the other two networks because some players are simultaneously playing a BoS and a PC game, which have different mixed strategy solutions, forbidding completely uncorrelated outcomes for the -D-C-D- and D-D-C networks.  

\section{Discussion}\label{discussion}

The study of the equilibria in the three types of networks considered here is naturally not exhausted and we leave this to future work. We have restricted ourselves to the class $2$ equilibria, and so looking at the class $3$ equilibria might also show interesting results. Furthermore, our assignments of $J$ and $J'$ and $B$ and $B'$ can be changed to reveal other network properties. For instance, in networks D-C-D and -D-C-D- we cannot distinguish if the differences arise due to a lack of games played between players $1$ and $3$ or due to the existence of a PC game. To this effect, other parametrizations are possible, such as labeling the interaction strength based on the existence of a link. The relationships between the magnetic fields could also be different, by assigning a different magnetic field depending on the number and type of game that is being played. There is evidence for this parametrization in the study of Broere \textit{et al.}\cite{broere2020}. 

In the context of other results in the literature, Hernandez \textit{et al.}\cite{hernandez2013heterogeneous} find that by reducing the knowledge of the players to a local level, for example to their own degree of connectedness, the set of Nash equilibria is limited when compared to having global knowledge, as there are less parameters to fine-tune, in line with the results of Galeotti \textit{et al.}\cite{galeotti2010network}. This provides support for the use of best response update rules as a good compromise between limited computational power and convergence to a Nash equilibrium, which happens as the local interactions between the players propagate to the entire network. Limited to this knowledge and probabilistic information about the degree distribution of the network, the expected equilibrium is that all players play D in the -D-C-D- and D-D-C networks, since the C node can have a high degree but is a minority, while for the D-C-D the expected equilibrium is that all play C, because this player is at the hub of the network\cite{hernandez2013heterogeneous}. In our model, the players have indirect access to global knowledge, via the three-player probability distribution, and therefore the number of nodes with a D preference has a large weight, which we can see as the bulk of the equilibria values happen for $B>0$. However, this is not deterministic and an entire range of initial correlations for each network can incorporate information that reflects a wide variety of final player configurations. In fact, our results show evidence that by quantifying the local degree of correlation between direct opponents in relation to a global probability distribution, it is possible to quantify the correlations between indirect players, which is the signature trait of the propagation of information inside the network. Extending our analysis to independent two-player correlations constitutes another direction of further research.

While we mostly study the three-player networks using an interaction strength that is shared between all pairs, imposing different external magnetic fields introduces different playing incentives. The results could then be understood in terms of spin frustration, where one of the spin particles has competing incentives to minimize its energy. Applying the techniques used for spin frustration on a triangular latice to the three-player network could unearth further symmetries that, if generalizable, would yield a powerful tool for the simulations of Nash equilibria. 

\section{Conclusion}\label{conclusion}

Asymmetric games on networks, played between agents with conflicting preferences, account for several real-life problems. Games on networks have been studied mostly numerically, where an update strategy dependent on the preferences or outcomes of neighboring players dictates their actions, and is applied repeatedly until convergence is reached. Through this process, the actions of one player reach beyond her direct opponents, which generates underlying correlations in the outcomes of all players. Describing the final configurations of networks with correlations using an Ising model offers a new way to study how these update rules lead to convergence in numerical simulations. The application of the slope analysis to the study of Nash equilibria in the presence of correlations on a subset of networks with three players provides an analytical framework to study which correlations can generate equilibria on a network, as a function of different configurations.

Our main finding is that the asymptotic behavior for large magnetic fields is not only common to all networks, but also contains information about the payoff structure, without any restriction imposed on the initial correlations \textit{a priori}. This is important because it hints at a universal behavior possibly extendable to larger networks, thus lifting the necessity to run expensive calculations.  However, further research is still needed to assess whether larger networks can be described using these three-player settings as building blocks, most immediately those containing nodes with only one or two degrees of connectedness. For smaller values of parameters, the network details become increasingly relevant. Another important result is the existence of regions that are out of equilibrium and the way in which they renormalize to an equilibrium, which allows us to analytically determine how the players must respond to a certain set of initial correlations if they want to reach a stable configuration other than an uncorrelated solution. 

While further work is necessary to establish the direct correspondence between the correlations used here and the update strategies in numerical simulations, we studied a number of parameters that allow for equilibria when correlations are involved. In particular, with our method we can quantify how these depend on the preferences of the players and games that they subsequently play with each other. Using the mapping to the Ising parameters, we were able to leverage the symmetries of different strategies and compare equilibria between different networks. We demonstrate in our study that for all types of networks there is a large phase space that accomodates correlations in equilibrium. This is consistent with the results of the ring-network simulations, where complex, but still Ising-like, correlations were found in the equilibrium configurations.

Correlations in games on networks are, beyond an artifice of numerical simulations, also evident in real life conflicts, as oftentimes a shared signal is used to reach convergence, even though the agents cannot communicate directly\cite{banks1992battle}. For this reason, quantifying the relationship between the many variables of a game, a network configuration, and the underlying correlations can be of great societal value. With the knowledge of the microscopic Ising parameters that govern the correlations, we can now use the appropriate statistical physics tools to observe and understand the resulting macroscopic behavior. As we arrive at a wide range of Ising models to tackle this problem, on which the advanced methods of statistical physics have not yet been applied, our method to study games on networks can in its turn stimulate new developments in statistical physics.

\begin{acknowledgments}

We thank Kevin Peters for his early work on the introduction of correlations in the BoS game. We also thank Vincent Buskens for helpful discussions that improved to our understanding of update strategies in simulations. This work is part of the D-ITP consortium, a program of the Netherlands Organization for Scientific Research (NWO) that is funded by the Dutch Ministry of Education, Culture and Science (OCW). This work is also supported by the UU Complex Systems Fund, with special thanks to Peter Koeze.

\end{acknowledgments}

\appendix
\section{Game Theory}

In this appendix we lay out a formal game theoretical treatment of the correlated games, for $n$ players. We start by defining strategic-form games, and their Nash equilibria solutions. Then we expand these games with correlations and define the correlated equilibrium. Finally, we introduce the response strategies that generate a renormalized set of correlations, and the conditions the guarantee that this is in correlated equilibrium. These conditions can be expressed in terms of the coefficients of the payoff function, which are used to evaluate which response strategies represent equilibria.

\subsection{Strategic form games}

A strategic form game is defined by three elements: the finite set $\mathcal{I}$ of $n$ players, with $\mathcal{I}=\{1,2,...,n\}$; the pure-strategy space $S_i$ for each player $i \in \mathcal{I}$, representing the plays that each player has available; and the payoff functions $u^i_{\mu_i,\mu_{-i}}$, denoting the gain of player $i$ when playing $\mu_i \in S_i$ and the other players, denoted by $-i$, play $\mu_{-i} \in S_{-i}$. Besides the pure strategies, the players can play a mixed strategy, in which player $i$ plays the pure strategy $\mu_i$ with probability $\sigma_i(\mu_i)$. A pure strategy is a particular case of a mixed strategy, that assigns probability $1$ to a certain element of the pure-strategy space.

The players do not know with certainty what their opponents will play, so the rational player has to consider all possible moves. Taking this into account, the Nash equilibrium guarantees that each player chooses a strategy from which they do not want to deviate. A mixed strategy profile $\sigma_i^*$ is the Nash equilibrium if for all players $i$ we have that their average payoff obeys 

\begin{equation} \label{weaker}
\langle u^i_{\sigma_i^*,\sigma_{-i}^*} \rangle \geq \langle u^i_{\mu_i,\sigma_{-i}^*} \rangle.
\end{equation} 
for all $s_i \in S_i$.
If the inequality is strict, a pure-strategy Nash equilibrium ensues. 

A game with two players is characterized by $\mathcal{I}={1,2}$, $S_i=\{C,D\}$, and payoff $u^i_{\mu_i,\mu_{-i}}$, in the case of the BoS and PC games as given in the entries of tables \ref{payoffs3}. That both players adopting the same strategy are pure Nash equilibria can be shown by the validity of Eq. (\ref{weaker}) with the corresponding payoff values.

\subsection{Correlated Games}

A correlation device is a probability distribution $p$ over the space of all possible joint outcomes $\Omega$ that informs each player of what they should play in order to reach a certain outcome $\omega \in \Omega$. For $n$ players, $\omega=\mu^\omega_1 \cdots \mu^\omega_i \cdots \mu^\omega_n$, where $\mu^\omega_i$ is the partial information $H_i(\omega)$ received by player $i$ that depends on the probability that the correlation device assigns to $\omega$. The situation is a correlated equilibrium if no agent can improve their expected outcome by unilaterally deviating from the correlation device's instructions. A strategy profile $(\mu^{\omega}_1,...,\mu^{\omega}_i,..,\mu^{\omega}_n)$ is a correlated equilibrium\cite{fudenberg1991game} if, for every player $i$,
        \begin{equation}
            \sum_{\omega\in\Omega}p_\omega u^i_{\mu^{\omega}_i,\mu^{\omega}_{-i}}\geq \sum_{\omega\in\Omega}p_\omega u^i_{\mu_i,\mu^{\omega}_{-i}}, 
        \end{equation} with $\mu_i$ any other strategy adopted by player $i$ other than the one recommended by the correlation device  $\mu^\omega_i$.
The correlations are introduced for the two-player BoS and PC games with the final state space given by $\Omega = \{ CC, CD, DC, DD\}$, and the correlation device described by the probabilities $p_{\mu_1,\mu_2}$, with $\mu_1 \in S_1$ and $\mu_2 \in S_2$. Going to the three-player setting, the possible outcomes are of the form $\mu_1 \mu_2 \mu_3$ and the correlations over them are given by $p_{\mu_1 \mu_2 \mu_3}$.

\subsubsection{Response Probabilities}\label{responsestrategies}

Response strategies give agents the freedom to choose whether to follow the instructions of the correlation device as their own strategy\cite{correia2019nash}. More concretely, a set of response probabilities $P^i_{F\mu^\omega_i}$ are added to the previous framework and it directly follows that the probability that advice $\mu_i^\omega$ is not followed by player $i$ equals $P^i_{NF\mu_i^\omega}=1-P^i_{F\mu_i^\omega}$. For $n$ players, the response probabilities induce a mapping from the initial probabilities $p_{\omega'}$, with $\omega'=\mu^{\omega'}_1 \cdots \mu^{\omega'}_i \cdots \mu^{\omega'}_n$ to the \textit{renormalized} probabilities $p^R_{\omega}$. Abreviating the notation from $\mu^\omega_i$ to $\mu_i$ and $\mu^{\omega'}_i$ to $\mu'_i$, and defining the response probabilities with respect to the transition probabilities $P^i_{\mu_i\leftarrow\mu_i'}$ as
\begin{equation}
    P^i_{\mu_i\leftarrow\mu_i'}=\delta_{\mu_i\mu_i'}P^i_{F\mu'_i}+(1-\delta_{\mu_i\mu_i'})P^i_{NF{\mu'_i}},
\end{equation} the renormalized correlation probabilities are given by as
\begin{equation}
    p^R_{\omega}=\sum_{\omega'} p_{\omega'} \prod_i^n  P^i_{\mu_i\leftarrow\mu_i'}.
\end{equation} 
The new strategy space where the actions are to follow or to not follow the instructions of the correlation device, the response strategies, define a new game, the \textit{correlated game}. For this game, the expected outcome for agent $i$ is given by
\begin{equation} \label{expectedoutcome_def}
    \langle u^i \rangle = \sum_{\mu,\nu}p^R_{\omega}u^i_{\omega}
    .
\end{equation}
An equilibrium in this framework is a situation where neither agent can improve its own expected payoff by changing its own response probabilities, given the strategy of the other agent. More formally, given the correlation device $p$ and that player $i$ received information $\mu'$, an $n$-agent response strategy profile $(P^{i*}_{F\mu_i},P^{-i*}_{F\mu_{-i}})$ is an equilibrium if for all $i$ and for all $P^i_{\mu_i\leftarrow\mu_i'}$ the following condition is true:
        \begin{align}
            &\sum_{\omega,\{\omega'\backslash \mu'_i\}} u^i_{\omega} p_{\omega'} P^{i*}_{\mu_{i}\leftarrow\mu'_{i}} \prod_{-i}^n  P^{-i*}_{\mu_{-i}\leftarrow\mu'_{-i}} \nonumber \\ \geq& \sum_{\omega,\{\omega'\backslash \mu'_i\}} u^i_{\omega} p_{\omega'} P^{i}_{\mu_{i}\leftarrow\mu'_{i}} \prod_{-i}^n  P^{-i*}_{\mu_{-i}\leftarrow\mu'_{-i}}
            ,
        \end{align} with $\{\omega'\backslash \mu'_i\}$ the set of pure strategies suggested to each player by the correlation device in order to achieve state $\omega'$ except the one suggested to player $i$.

We denote a specific set of response strategies by  $\left(P^1_{FC},P^1_{FD},...,P^{n}_{FC},P^n_{FD}\right)$, not necessarily in equilibrium, bearing in mind that it is always relative to a certain initial set of probabilities, which should be clear from the context. If we want to refer to the strategy of a particular player $i$ in contrast with those of its adversaries $-i$, we write the response strategy as $ \left(P^{i}_{F\mu_i}, \left\{P^{-i}_{F\mu_{-i}}\right\} \right)$. The expected outcome, as given in eq.  (\ref{expectedoutcome_def}), is an expression which is linear in the response strategy of the agents. We can rewrite that expression with respect to the \textit{slope coefficients} $C^i_{\mu_i}$ as functions of the initial correlation device and of the response strategies of the adversaries:

\begin{equation}\label{slope coefficients}
    \langle u^i \rangle = C^i\left(\left\{P^{-i}_{F\mu_{-i}}\right\}, p_\omega \right) + \sum_{\mu_i} C^i_{\mu_i}\left(\left\{P^{-i}_{F\mu_{-i}}\right\}, p_\omega\right) P^{i}_{F\mu_i}.
\end{equation}
 Given the response strategies of the other agents, the slope coefficients determine whether a strategy for agent $i$ is a Nash equilibrium with respect to their own response strategies, which means that they cannot improve their outcome by unilateray deviating from the equilibrium solution. This is the case in three situations: 
\begin{enumerate}
    \item {$C^i_{\mu_i}>0$, and $P^i_{F\mu_i}=1$;}
    \item {$C^i_{\mu_i}<0$, and $P^i_{F\mu_i}=0;$}
    \item {$C^i_{\mu_i}=0$.} 
\end{enumerate}
The first two cases give pure equilibria in the space of response strategies, while the last case case gives a mixed equilibrium. For each combination of equilibrium slope conditions it can be calculated whether there are correlation devices for which that strategy results in a Nash equilibrium. To guarantee that a strategy is an equilibrium for all players simultaneously, the parameters of players $-i$ that are used in the slope coefficients of player $i$ should have values that are \textit{consistent} with the signs of the slope coefficients of those parameters when calculating the expected payoffs of players $-i$.

\section{Combinatorics of the Response Strategies}\label{combinatorics}

A range of $81$ response strategies of the form $\left(P^1_{FC},P^1_{FD},P^2_{FC},P^2_{FD}\right)$ are possible solutions of the two-player Battle of the Sexes game. However, some strategies yield slope conditions that are conflicting or unsatisfiable, eliminating them from the outset. We write a slope associated to the action player $i$ playing $\mu_i$ as $C^i_{\mu_i}\left(P^{-i}_C,P^{-i}_D\right)$, with a dependence on the response strategies $P^{-i}_{\mu_{-i}}$ of its opponent $-i$. In this appendix, we evaluate systematically the slopes and the conditions that the response strategies impose on them, ordered by the total number of slopes that are equal to zero. The imposition of this equality on a slopes results in an equilibrium probability $P^{i*}_{\mu_i}$ that lies between $0$ and $1$.

\subsection{No slope condition is equal to zero}

If the slope conditions are not zero, then they can only be positive or negative, which means that the probabilities can only take the values of either $0$ or $1$. This gives a total of $2^4=16$ possibilities. Of these, we will exclude the $8$ that have a majority of one of the values, such as in $(1,1,1,0)$, and additionally the strategies $(1,0,0,1)$ and $(0,1,1,0)$.

In the first case, the equilibrium conditions will generate conflicting conditions. Taking the example strategy, the condition related with the probability of player $1$ playing D takes the value $$C^1_D(1,0)=-p_{DD}-p_{DC},$$ which is negative. However, for its own consistency, player $1$ should always follow an instruction to play D, such that $C^1_D>0$. This is inconsistent with the structure of this game, and therefore the response strategy $(1,1,1,0)$ cannot be an equilibrium. A similar reasoning allows us to exclude the remaining $7$ strategies.

In the second case, strategy $(0,1,1,0)$ requires exactly the same condition for $C^1_D$ as in the previous example, and hence yields the same contradiction. Because strategy $(1,0,0,1)$ contains a similar contradiction as strategy $(0,0,0,1)$, it will also be excluded.

The fact that all probabilities sum to $1$ determines that the number of degrees of freedom of the initial correlation probabilities is at most three. The slope conditions of the strategies without zero-valued conditions thus define a four-faced volume on the space of the initial probabilities, where they correspond to equilibria. In our analysis of Sec. \ref{twoplayergames} we further remove one degree of freedom by imposing that $p_{CD}=p_{DC}$, which reduces the analyzed region from a volume to a plane. This remark will be relevant to the analysis of the next type of strategy.

\subsection{One slope condition is equal to zero}

When one of the four slopes has a value of zero, the remaining three slopes can be either positive or negative. Therefore, there are $4\times 2^3=32$ such strategies. Comparing with the previous strategies, the slope conditions have now an equality in what was before the inequality condition that bounded the equilibrium volume. It remains to see what happens at the symmetric section.

Let us consider the response strategy $(P^{1*}_{FC},0,0,0)$, with the slope conditions

\begin{align}
\left\{\begin{matrix}
&C^1_C(0,0) =-sp_{CC}+p_{CD}= 0\\
&C^1_D(0,0) =-p_{DD}+sp_{DC}<0\\
&C^2_C(P^{1*}_{FC},0) =p_{CC}\left((s+ 1)P^{1*}_{FC}-1\right) +sp_{DC}<0\\
&C^2_D(P^{1*}_{FC},0) =-sp_{DD}-p_{CD}\left((s+ 1)P^{1*}_{FC}-1\right)<0
\end{matrix}\right. .
\end{align} The first condition imposes that $p_{CC}=p_{CD}/s$, which using the symmetry condition $p_{CD}=\left(1-p_{CC} - p_{DD} \right)/2$ translates into the equality $p_{DD}=1-\left({s/2}+1 \right)p_{CC}$. On the plane represented in Fig. \ref{fig:PC2_combined}, this line is the boundary shared between regions A and C, B and D, and G and J. The other three conditions on the symmetric place correspond, respectively, to

\begin{align}
 &   p_{DD}> \frac{s(1-p_{CC})}{2+s},\label{1stcons} \\ 
 &   P^{1*}_C < (1-s), \label{2ndcons} \\  
&  P^{1*}_C > \frac{1-\frac{p_{DD}}{p_{CC}}}{s+1}.   \label{3rdcons}
\end{align} Condition (\ref{1stcons}) relates to the area above the line that is the boundary between regions A and B, C and D, and H and K of Fig. \ref{fig:PC2_combined}. Condition (\ref{3rdcons}) is only consistent with condition (\ref{2ndcons}) as long as $p_{DD}> s^2 p_{CC}$, which corresponds to the area above a line going through the origin and the intersection point of regions C, F, H, K, I and D. Therefore, we can conclude that this strategy is an equilibrium in the line segment that starts at $p_{DD}$ and ends at the intersection point of regions A, C, D and B. 

Outside of the symmetric section, we see that the condition that is being equated to zero corresponds to one of the faces of the solid that would correspond to the equilibrium region of $(0,0,0,0)$ or $(1,0,0,0)$. As we have already seen that the second strategy cannot generate an equilibrium, we can conclude that strategy $(P^{1*}_C,0,0,0)$ is, by continuity, an equilibrium at the boundary of the solid where $(0,0,0,0)$ is an equilibrium. However, we cannot say the same at the symmetric section, as the line segment where we found this strategy to be equilibrium is not adjacent to the areas K and L, differently from what would happen, for instance, to strategy $(0, P^{1*}_D,0,0)$, where the equality would contain, but not be limited to, the lower boundary of region K.

Similarly for the remaining $31$ strategies, we find that they will be boundaries of other equilibrium regions when we consider the full space of initial correlations. On the symmetric plane, however, they may or may not connect to the boundaries of the equilibrium regions. Due to their liminal character, we choose not to analyze these strategies, arguing that if, on the one hand, they correspond to boundaries, then the properties of the bulk must extend to them, and that if, on the other hand, they are not so connected, then they consist of only a line segment and its analysis is not relevant in the scope of this work.

\subsection{Two slope conditions are equal to zero}

The possible combinations of two positive- or negative-valued slopes gives a total of $\left((2!\times 2!)/4\right) \times 2^2=24$ response strategies of this kind. Of these, only the $16$ strategies in which each player has both a condition with a zero-valued slope and with a non-zero-valued slope  will be equilibria, as for instance $(P^{1*}_{FC},0,1,P^{2*}_{FD})$. 

To realize that the remaining $8$ strategies, where each player has only one type of slope, cannot be equilibria, consider the response strategy $(P^{1*}_{FC},P^{1*}_{FD},0,0)$. The full set of conditions is presented below:

\begin{align}
\left\{\begin{matrix}
C^1_C(0,0) &=-s p_{CC}+p_{CD}= 0\\
C^1_D(0,0) &=-p_{DD}+sp_{DC}= 0\\
C^2_C\left(P^{1*}_{FC}, P^{1*}_{FD}\right) &=p_{CC}\left((s+ 1)P^{1*}_{FC}-1\right)\\
&-p_{DC}\left((s+ 1)P^{1*}_{FD}-s\right)<0\\
C^{2}_D\left(P^{1*}_{FC}, P^{1*}_{FD}\right) &=p_{DD}\left((s+ 1)P^{1*}_{FD}-s\right)\\
&-p_{CD}\left((s+ 1)P^{1*}_{FC}-1\right)<0
\end{matrix}\right. .
\end{align} Substituting the first two equations into the last two conditions results in the following expressions:

\begin{align}
    &p_{CC}\left((s+ 1)P^{1*}_{FC}-1\right) < p_{DD}\left((s+ 1)P^{1*}_{FD}-s\right),\\
    & p_{CC}\left((s+ 1)P^{1*}_{FC}-1\right) > p_{DD}\left((s+ 1)P^{1*}_{FD}-s\right),
\end{align} which contradict each other. A similar reasoning shows that the other $7$ strategies cannot cannot have equilibrium solutions.

\subsection{Three slope conditions are equal to zero}

In this case, one of the four slopes can be either positive or negative, which translates in $4\times 2=8$ possible response strategies, but none of them can be an equilibrium. Take for example the response strategy $(P^{1*}_{FC},P^{1*}_{FD},P^{2*}_{FC},0)$. For player $1$ we obtain the following equilibrium conditions

\begin{align}
\left\{\begin{matrix}
C^1_C\left(P^{2*}_{FC},0\right) &=p_{CC}\left((s+ 1)P^{2*}_{FC}-1\right) +p_{CD} =0\\
C^{1}_D\left(P^{2*}_{FC},0\right) &= -p_{DD}-p_{DC}\left((s+ 1)P^{2*}_{FC}-s\right) =0
\end{matrix}\right. ,
\end{align} which can be solved to give the following simple condition on the initial probabilities

\begin{equation}\label{condition}
    p_{CD}p_{DC}=p_{CC}p_{DD}.
\end{equation} For player $2$ we have the conditions 

\begin{align}
\left\{\begin{matrix}
&C^2_C\left(P^{1*}_{FC}, P^{1*}_FD\right)  =p_{CC}\left((s+ 1)P^{1*}_{FC}-1\right)\\
&-p_{DC}\left((s+ 1)P^{1*}_{FD}-s\right)= 0\\
&C^2_D\left(P^{1*}_{FC}, P^{1*}_{FD}\right) = p_{DD}\left((s+ 1)P^{1*}_{FD}-s\right)\\
&-p_{CD}\left((s+ 1)P^{1*}_{FC}-1\right) <0.
\end{matrix}\right. .
\end{align} Using Eq.(\ref{condition}), these conditions reduce to

\begin{align}
    &\frac{p_{CC}}{p_{DC}}\left((s+ 1)P^{1*}_{FC}-1\right) = \left((s+ 1)P^{1*}_{FD}-s\right),\\
    &\frac{p_{CC}}{p_{DC}}\left((s+ 1)P^{1*}_{FC}-1\right)<\left((s+ 1)P^{1*}_{FD}-s\right).
\end{align} We see that these conditions are incompatible, which means that this response strategy cannot be an equilibrium anywhere on the set of initial correlations. Extending this reasoning to the remaining $7$ strategies yields equivalent results.

\subsection{All slope conditions are equal to zero}

There is only one such response strategy, $(P^{1*}_{FC},P^{1*}_{FD},P^{2*}_{FC},P^{2*}_{FD})$, which is always in equilibrium, as it corresponds to a system of four equations, given by the slope conditions, with four independent variables, the follow probabilities. This system has a unique solution that is ultimately independent of the initial probabilities, in which the response probabilities exactly reproduce the mixed strategy Nash equilibrium probabilities of the uncorrelated game.

\bibliography{apssamp}% Produces the bibliography via BibTeX.

\end{document}